\documentclass[aip,pof,reprint,amsmath,amssymb,floatfix]{revtex4-1}
\usepackage{graphicx,subfig,booktabs}

\usepackage[svgnames,table,hyperref]{xcolor} %
\definecolor{dblue}{rgb}{0 0.447 0.741}
\definecolor{dgreen}{rgb}{0.274 0.784 0.424}
\definecolor{dred}{rgb}{0.7 0 0.1}

\usepackage{dcolumn}%
\usepackage{bm}%

\usepackage[utf8]{inputenc}
\usepackage[T1]{fontenc}
\usepackage{mathptmx}
\usepackage{units}	%
\usepackage{morefloats}

\newcommand{\bi}{ \boldsymbol}%

\usepackage[colorlinks,linkcolor=dblue,citecolor=dgreen]{hyperref}	%

\begin{document}

\title{Synchronizing the helicity of Rayleigh-B\'enard convection by a tide-like electromagnetic forcing}

\author{Peter J\"ustel}
  \altaffiliation{Helmholtz-Zentrum Dresden - Rossendorf, Bautzner
	Landstr. 400, 01328 Dresden, Germany}

\author{Sebastian R\"ohrborn}
  \altaffiliation{Helmholtz-Zentrum Dresden - Rossendorf, Bautzner
	Landstr. 400, 01328 Dresden, Germany}

\author{Sven Eckert}
  \altaffiliation{Helmholtz-Zentrum Dresden - Rossendorf, Bautzner
    Landstr. 400, 01328 Dresden, Germany} 

 \author{Vladimir Galindo}
  \altaffiliation{Helmholtz-Zentrum Dresden - Rossendorf, Bautzner
    Landstr. 400, 01328 Dresden, Germany} 
    
 \author{Thomas  Gundrum}
  \altaffiliation{Helmholtz-Zentrum Dresden - Rossendorf, Bautzner
    Landstr. 400, 01328 Dresden, Germany}
    
\author{Rodion Stepanov}
  \altaffiliation{Institute of Continuous Media Mechanics,1 Acad. Korolyov str., 614013 Perm, Russia}

\author{Frank Stefani}
  \altaffiliation{Helmholtz-Zentrum Dresden - Rossendorf, Bautzner
	Landstr. 400, 01328 Dresden, Germany}

\date{\today}

\begin{abstract}
We present results on the synchronization of the helicity in a 
liquid-metal 
Rayleigh-B\'enard (RB) experiment under the influence of a tide-like  
electromagnetic forcing with azimuthal wavenumber $m=2$. We show that for 
a critical forcing strength the
typical Large Scale Circulation (LSC) in the cylindrical vessel of aspect ratio unity 
is entrained by the period of the tide-like forcing, leading to 
synchronized helicity oscillations with opposite signs in two half-spaces.
The obtained experimental results are consistent with and supported by numerical simulations. 
A similar entrainment mechanism for the helicity in the solar 
tachocline may be responsible for the astonishing 
synchronization of the solar dynamo by the 11.07-year 
triple 
synodic alignment cycle of the tidally dominant planets Venus, 
Earth and Jupiter.

\end{abstract}



\maketitle

\twocolumngrid %

\section{Introduction} \label{introduction}

Helicity, i.e. the scalar product of a vector field with 
its vector potential, plays a key role in various fields of 
hydrodynamics and plasma physics  \cite{Moffatt2018}. 
A point in case  is the decisive role
of {\it magnetic} helicity conservation for 
Taylor relaxation in fusion plasmas \cite{Taylor1986}.
A complementary example is the importance of
{\it kinetic} helicity for
magnetic field self-excitation in 
planets, stars, and galaxies \cite{Rincon2019,Tobias2021}, 
as recently confirmed in various liquid metal dynamo 
experiments \cite{Zamm2008}. 
In those cosmic bodies, and technical devices, 
helicity triggers 
the induction
of electrical currents in the direction of a prevailing 
magnetic field. This so-called $\alpha$ effect
is at the root of $\alpha^2$ dynamos, thought to be at work 
in the Earth's liquid core,
as well as of 
$\alpha-\Omega$ dynamos, responsible for 
the magnetic field generation in sun-like stars.

While contemporary solar dynamo theory \cite{Charboneau2020}
has been quite successful
in reproducing the main features of the solar magnetic
field, such as the typical time scale of the Schwabe cycle and
the shape of the butterfly diagram of sunspots, there remain some 
nagging doubts concerning the conceptional completeness of 
the models employed so far. One of the 
unresolved issues relates to the astonishing regularity of the
11-year Schwabe cycle which, notwithstanding some fluctuations,
shows statistical features of a clocked process 
which distinguishes it 
from a simple random walk process \cite{Dicke1978}. 
Meanwhile, we have 
remarkable observational evidence for the phase-stability of the
Schwabe cycle,
stemming  
both from sunspot and cosmogenic nuclides' data of 
the last millennium, pointing to a cycle length of 11.07 years, 
as well as from algae data of the early Holocene, showing a 
(barely distinguishable) period of 11.04 years 
\cite{Vos2004,AN2020}. 

Inspired by  previous work of Hung \cite{Hung2007}, 
Wilson \cite{Wilson2008,Wilson2013} and Scafetta 
\cite{Scafetta2012}, we 
have pursued the idea of a  
possible link between the Schwabe cycle and the 
11.07-year triple synodic cycle 
of the tidally dominant 
planets Venus, Earth and Jupiter
\cite{Weber2015,Stefani2016,Stefani2018,Stefani2019,Stefani2020,Stefani2021}. 
A promising rationale for such a connection was found in 
the tendency of the current-driven, kink-type ($m=1$) Tayler 
instability (TI) \cite{Tayler1973,Seilmayer2012} to undergo 
intrinsic {\it helicity oscillations} \cite{Weber2015}, which later were 
shown to be easily entrainable by a
tide-like perturbation with its typical 
$m=2$ azimuthal dependence \cite{Stefani2016}.
While, admittedly, the tidal forces exerted by the planets in the
solar tachocline are very weak \cite{Callebaut2012}, 
the helicity entrainment mechanism  employs them only as a 
catalyst to switch between left- and 
right-handed states of the pre-existing TI, leaving 
its very energy content
nearly unchanged.

Yet, the TI is just one candidate 
for the tides to act upon. A very similar synchronization 
mechanism might apply to  magneto-Rossby
waves at the  solar tachocline 
which are presently under intense investigation 
\cite{Dikpati2017,McIntosh2017,Tobias2017,Zaqarashvili2018,Zaqarashvili2021}.
Preliminary results suggest that realistic tidal forces
as exerted by planets
are capable of exciting
magneto-Rossby waves
with velocity amplitudes that could indeed be 
relevant for the solar dynamo \cite{Horstmann2022}. 

A related  question that presents itself concerns 
the possible influence of tidal forces
on convective motion \cite{Schumacher2020}.
While, due to  the huge differences 
between the tidal and inertial forces in the solar convection zone 
\cite{Callebaut2012,Charbonneau2022}, 
any noticeable effect in this region 
seems extremely unlikely, convection is 
at least an attractive candidate for an
experiment on helicity synchronization. 
In view of the significant 
challenges in carrying out liquid metal experiments 
on the very TI \cite{Seilmayer2012},
Rayleigh-B\'enard(RB) convection may  provide
an interesting {\it surrogate} with nearly identical 
topological features. Just as the TI,
the Large Scale Circulation (LSC) 
\cite{Krishnamurti1981,Sano1989,Takeshita1996,Cioni1997,Kadanoff2001,XiLamXia2004,BrownAhlers2006,Resagk2006,Xi2008} 
in an H/D aspect ratio unity vessel breaks 
spontaneously the 
axi-symmetry ($m=0$) 
of the underlying problem and develops 
an $m=1$ flywheel structure.
Secondary effects such as torsional \cite{Funfschilling2004}
and 
sloshing modes \cite{XiZhou2009,BrownAhlers2009},
reversals and even intermittent cessations \cite{BrownAhlers2006,Xi2008}
were experimentally studied with 
different working fluids, including 
water \cite{Krishnamurti1981,BrownAhlers2006,Xi2008}, 
silicon oil \cite{Krishnamurti1981}, helium-gas \cite{Sano1989}, 
air \cite{Resagk2006}, 
liquid mercury \cite{Takeshita1996,Cioni1997},
liquid sodium \cite{Khalilov2018}, and the 
eutectic alloy GaInSn \cite{Wondrak2018,Zuerner2019}.
Since the sloshing mode with its side-wise motion 
(transverse to the primary LSC vortex) is also connected 
with a helicity oscillation, the interaction of the $m=1$ 
LSC with some $m=2$ tide-like perturbation seems attractive   
for a paradigmatic experimental verification of the 
more generic helicity synchronization mechanism.

With this goal in mind, we have recently 
proposed \cite{StepanovStefani2019}, 
and later confirmed  \cite{POF2020},
how an electromagnetic tide-like forcing 
can be 
realized in a liquid-metal filled cylindrical cell of 
aspect ratio unity.
For that purpose, we utilized two 
coils (located opposite each other) of the more
versatile MULTIMAG system \cite{Pal2009} (which, in principle, 
allows for arbitrary superpositions of axial, rotating and 
travelling magnetic fields). Feeding these two coils with
AC currents (with typical frequencies if 25-100 Hz) we were able to
produce four-roll structures with an approximate 
$m=2$ symmetry\cite{POF2020}. 

As a sequel to those preliminaries\cite{StepanovStefani2019,POF2020}, 
this paper is dedicated to the
investigation of the
interaction of a tide-like electro-magnetic forcing 
with the single-roll LSC as it typically develops in cylindrical RB 
convection with aspect ratio unity.
Our goal is to understand if, how, and under which conditions, the $m=2$ forcing
leads to a synchronization of the sidewise motion of the LSC, and the helicity
connected with that.
In the next section we will recall the experimental set-up and 
the numerical methods for simulating the experiment. 
Then we will present the main experimental findings and compare them
with numerical results. The paper concludes with a summary of our 
findings and a discussion of possible lessons to be 
learned for the original solar dynamo problem. 

\section{Experimental and numerical setups} \label{experiment}

For the sake of completeness, in this section 
we recall the set-up of the experiment and the 
numerical solver used for its simulation. More details
can be found in previous work
\cite{POF2020,Roehrborn2022a,Roehrborn2022b}.

\subsection{Experimental setup}
\begin{figure}
	\centering
	\includegraphics[width=0.35\textwidth]{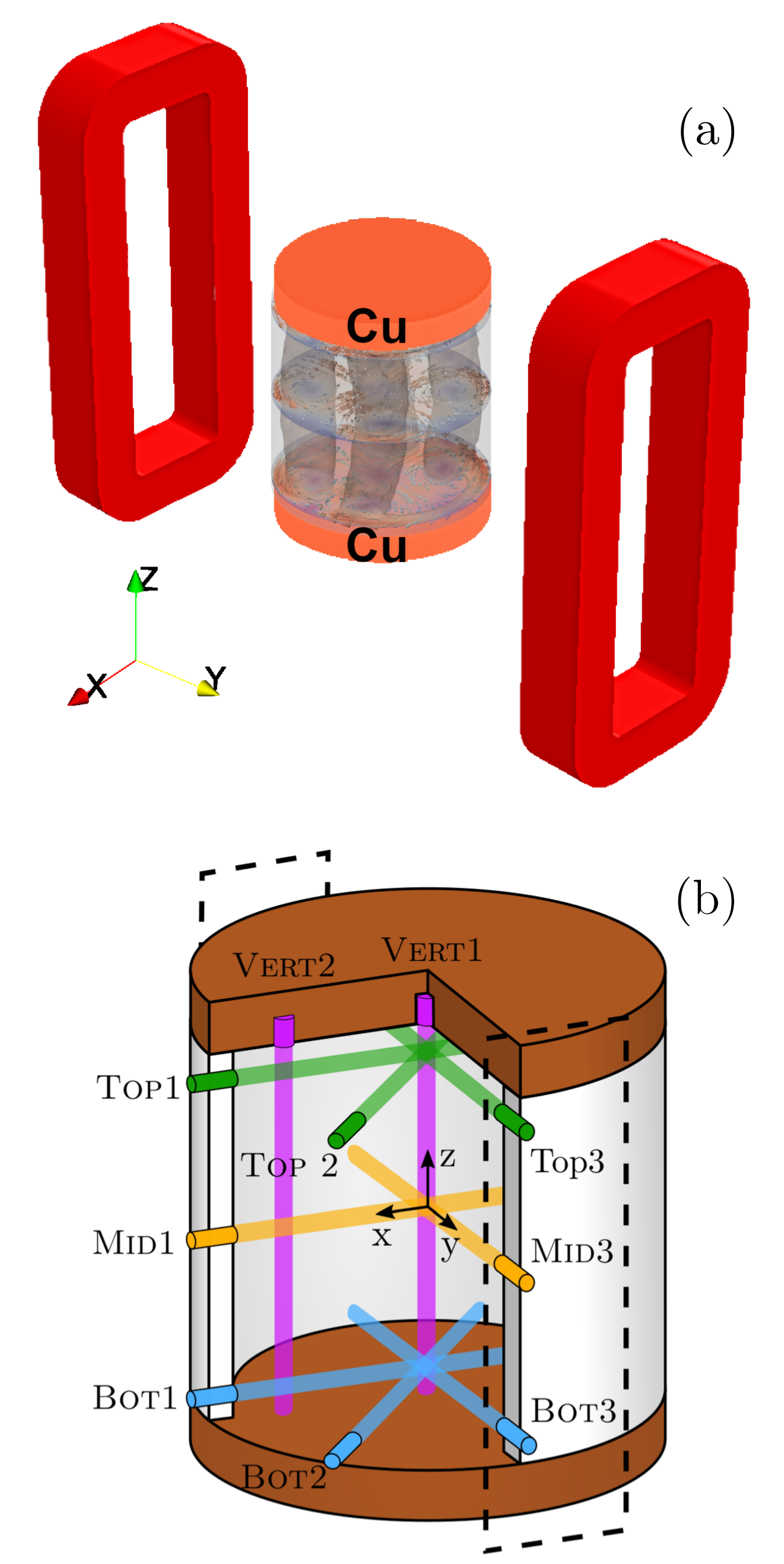}
	\caption{Schematic setup of the experiment. (a) Two coils (red), fed
	by an AC current with $25$\,Hz,
	produce a tide-like $m=2$ forcing in the cylindrical container filled with 
	GaInSn. In the case without RB flow, the tide-like force generates
	4 rolls which are also shown \cite{POF2020}. (b) 
	Configuration of the UDV sensors and the measurement paths. 
	Here, the arrangement of the excitation coils lying on the y-axis is 
	only indicated by the dashed lines.}
	\label{setup}
\end{figure}

The experiments are performed in a cylindrical container 
of aspect ratio $\Gamma=D/H=1$, with the height $H$ and the
diameter $D=2R$ both being equal to 180\,mm (see 
Fig.\,\ref{setup}).
As working fluid we use the liquid metal alloy GaInSn with 
the following physical parameters (at $20^{\circ}$\,C, see 
\cite{Plevachuk2015}): density $\rho=6350$\,kg/m$^3$, 
kinematic viscosity $\nu=3.44 \times 10^{-7}$\,m$^2$/s, 
thermal diffusivity $\kappa =1.19 \times 10^{-5}$\,m$^2$/s,
electrical conductivity $\sigma=3.27\times 10^6$ ($\Omega$ m)$^{-1}$.
From the latter values we infer a Prandtl number
$Pr=\nu/\kappa=0.029$ and a magnetic Prandtl number 
$Pm=\mu_0 \sigma \nu=1.40 \times 10^{-6}$.

The sidewalls, made of polyether ether ketone (PEEK), are considered
thermally and electrically insulating.  On top and 
bottom, the cylinder is bounded by two 
uncoated copper plates with \unit[220]{mm} diameter and a thickness 
of \unit[25]{mm}. During the RB experiments the bottom plate is electrically heated
while the top plate is watercooled. If not otherwise stated we
work with a temperature difference of \unit[5]{K} which is ensured by 
a thermostat. This temperature difference corresponds to a 
Rayleigh number $Ra=\alpha g H^3 \Delta T /\nu \kappa=1.03 \times 10^{-7}$ with thermal expansion coefficient $\alpha$ and gravitational acceleration $g$. 
The turnover time $t_{to} = L/\overline{u}_{LSC}$ with $L$ being the LSC path length for this state is estimate to be around \unit[50]{s}. 
From the electromagnetic point of view 
both copper plates are considered to be in perfect electrical 
contact with the GaInSn. Due to the large ratio (appr. factor 18) 
between the
thermal conductivities of copper and GaInSn and a small size ratio resulting in a low Biot Number $\ll 1$\cite{Schindler2022}, the 
temperature distribution at the interface between the two materials 
is considered  homogeneous.

Just as in \cite{POF2020}, the tide-like $m=2$ forcing 
is generated by AC-currents through two rectangular, 
stretched coils which are situated on opposite sides of the cylinder 
as delineated in  Fig.\,\ref{setup}a. Each has 80 turns, 10 pancake layers,  
a total inner height of \unit[350]{mm} and a mean distance between the 
long leg and the x-z centre plane of \unit[145]{mm}\cite{Pal2009}). The distance between the central pancake layer and the cylinder centre is \unit[285]{mm}.
When a current is applied to the coils a magnetic field is generated 
which is symmetrical to the centre plane. A tunable AC power supply 
is used to create an alternating current with defined amplitude and 
frequency in the coils. Throughout this paper, this frequency will 
be fixed to
\unit[25]{Hz} which was optimised previously
\cite{POF2020}. The polarity of the coils is assigned in a way 
that the magnetic field is concordant in both solenoids. This configuration 
has turned out advantageous for generating the desired flow 
structure \cite{StepanovStefani2019}.  

Although the cylinder is also equipped with thermo-couples, in this 
paper we exclusively rely on the velocity signals measured 
by \unit[8]{MHz} Ultrasound Doppler Velocimetry (UDV) sensors, whose
configuration around the vessel is illustrated in 
Fig.\,\ref{setup}b. The sampling period for all the sensors is about \unit[2.45]{s} for the experimental data and \unit[1]{s} (plain RB) and \unit[2]{s} for the numerical data.
Referring to the bottom interface between copper and GaInSn, 
the measurement planes ``Bot'', ``Mid'' and ``Top'' are located at 
heights of 10, 90 and \unit[170]{mm}, respectively. At each height, 
two sensors called ``1'' and ``3'' are placed with an angle 
of $\pi/2$ between them (the additional sensors  ``2'' are not 
utilized in the following). Two 
further UDV sensors for measuring the vertical flow component are 
placed in the top copper plate at radial positions 
$r/R=0$ and $r/R=0.8$. Of those, we will only utilize data 
of  ``Vert2'', situated close to the lateral boundary.
A more detailed description of the cell can be found in 
\cite{Zuerner2019}.

\subsection{Numerical scheme}

The flow in the cell is computed, using OpenFOAM 6 \cite{openfoam}, by solving the 
incompressible Navier-Stokes equation and
the continuity equation
\begin{equation} \label{ns}
\rho \frac{\partial \bi u}{\partial t} + \rho (\bi u \cdot \nabla ) \bi u
= - \nabla p + \rho \nu \nabla^2 \bi u + \bi f \, ,
\end{equation}
\begin{equation} \label{ce}
\nabla \cdot \bi u=0 \, ,
\end{equation}
wherein the force $\bi f$ is a sum of the electromagnetic force 
generated by the AC current in the coils and 
the buoyancy force due to the imposed temperature difference between the
lower and upper copper plates:
\begin{equation} \label{force}
\bi f = M(t) \bi f_{EM} + \bi f_{buoyancy} \,.
\end{equation}  

The thermal convection is modelled by the Boussinesq
approximation which provides the buoyant force $\bi f_{buoyancy}$ 
to the Navier-Stokes equation.
Since the resulting flow speeds are low (a few mm/s), the corresponding 
small magnetic
Reynolds numbers ($Rm \sim 10^{-3}...10^{-2}$ ) allow us to decouple 
the magnetic field generation from the flow
calculation. 

We therefore pre-computed the Lorentz force
\begin{equation} \label{force1}
\bi f_{EM}= \bi j \times  \bi B
\end{equation} 
by solving Maxwell's equations in Opera 1.7\cite{opera2018} and added 
it as a vector field to the Navier-Stokes
equation. To emulate a {\it time-dependent} tide-like
forcing, this time-constant part $\bi f_{EM}$ of the Lorentz force  
is amplitude-modulated by the factor
\begin{equation} \label{modulation}
M(t)=\sin^2(\pi f_{mod} t)
\end{equation}
wherein the frequency $f_{mod}$ is chosen 
close to the {\it natural} frequency $f_{slosh}$ of the 
sloshing motion of the LSC (which is displayed below).
To achieve a positive-defined sinusoidal modulation, i.e.
$0<M(t)<1$, the sine function is squared and the
natural frequency is halved.
Figure\,7 in \cite{POF2020} 
shows the spatial distribution of $\bi f_{EM}$
at the maximum $M(t)=1$.

The mesh utilized for the simulations consists of hexahedral cells 
with contracted cells at the walls where the
no-slip condition ${\bf u}=0$ is implemented. For all numerical 
simulations of the combination of RB convection and
tidal forcing, we used the results from a pure RB flow 
after 3000 seconds as starting point.\\ %

\section{Results}

\begin{figure*}
	\centering
	\includegraphics[width=\textwidth]{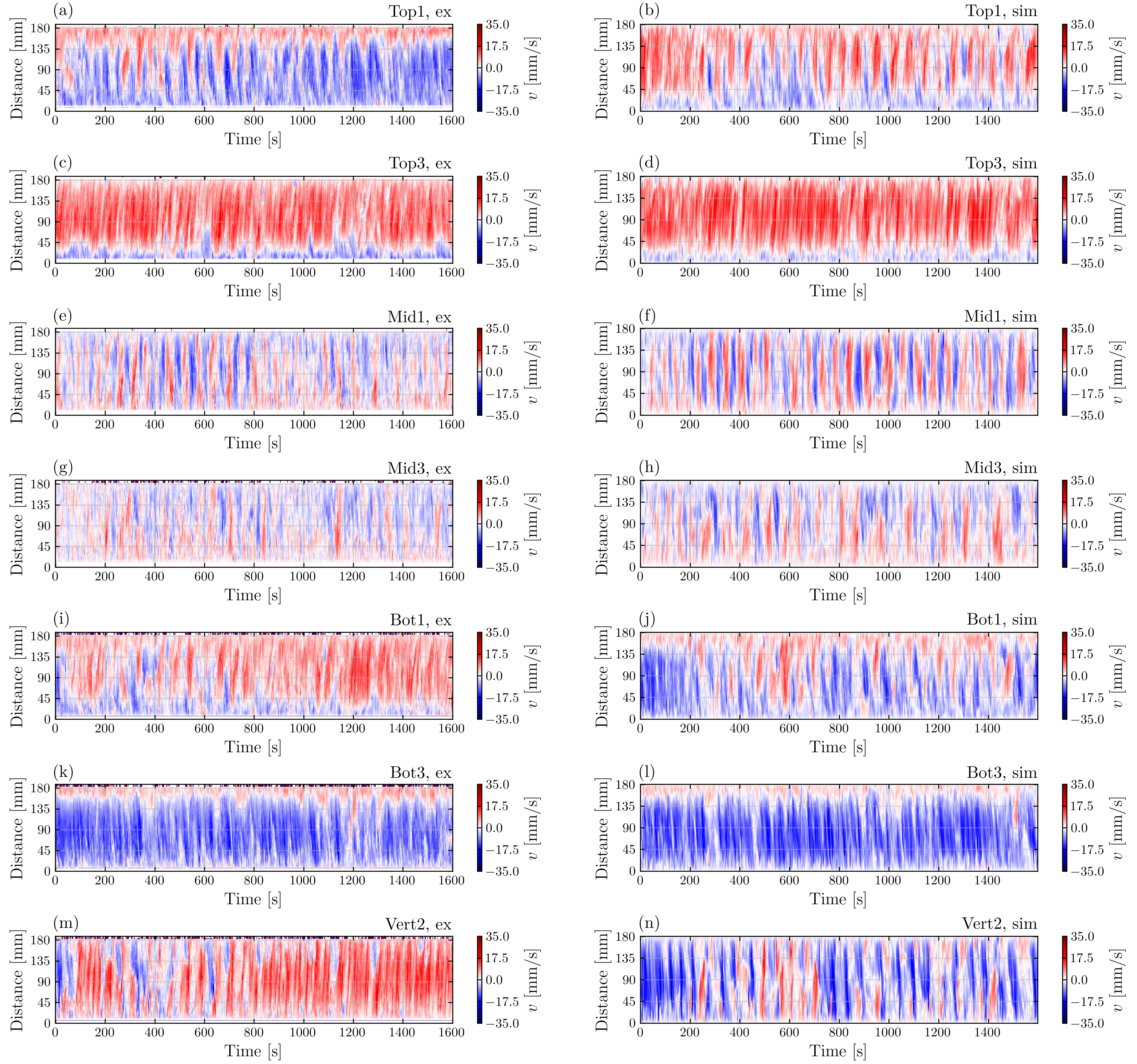}
	\caption{Contour plots of the velocities along the sensor beamlines 
		for a coil current of \unit[0]{A} (pure RB convection). 
		The left column shows measurement data for seven selected UDV sensors, while the right column 
		shows the corresponding virtual sensor data extracted from the simulation.\label{resultsContour0A}}
\end{figure*}

\begin{figure*}
	\centering
	\includegraphics[width=\textwidth]{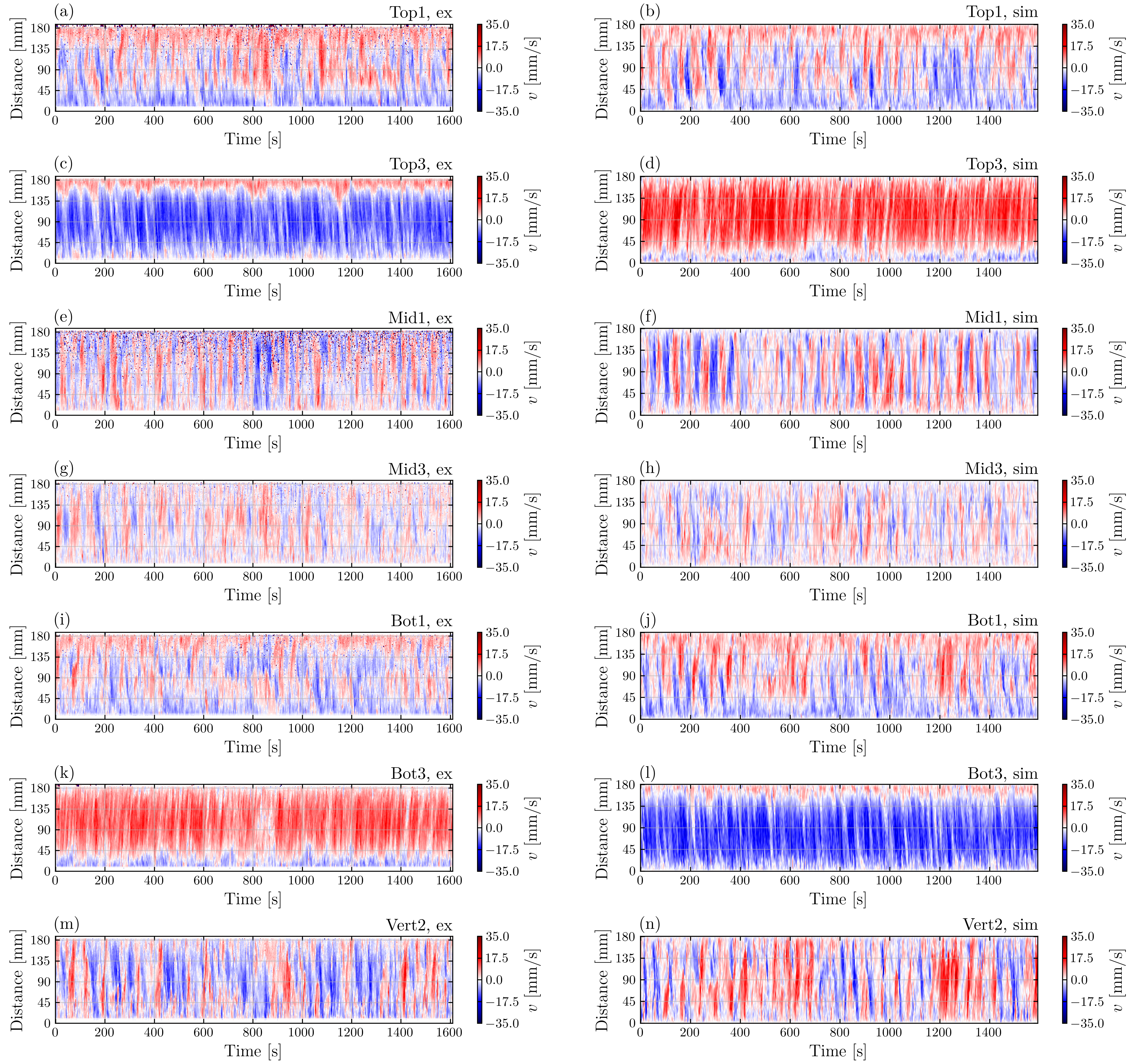}
	
	\caption{Same as Fig.\,\ref{resultsContour0A}, but for a coil current of \unit[12.5]{A}.
		\label{resultsContour12k5A}}
\end{figure*}

\begin{figure*}
	\centering
	\includegraphics[width=\textwidth]{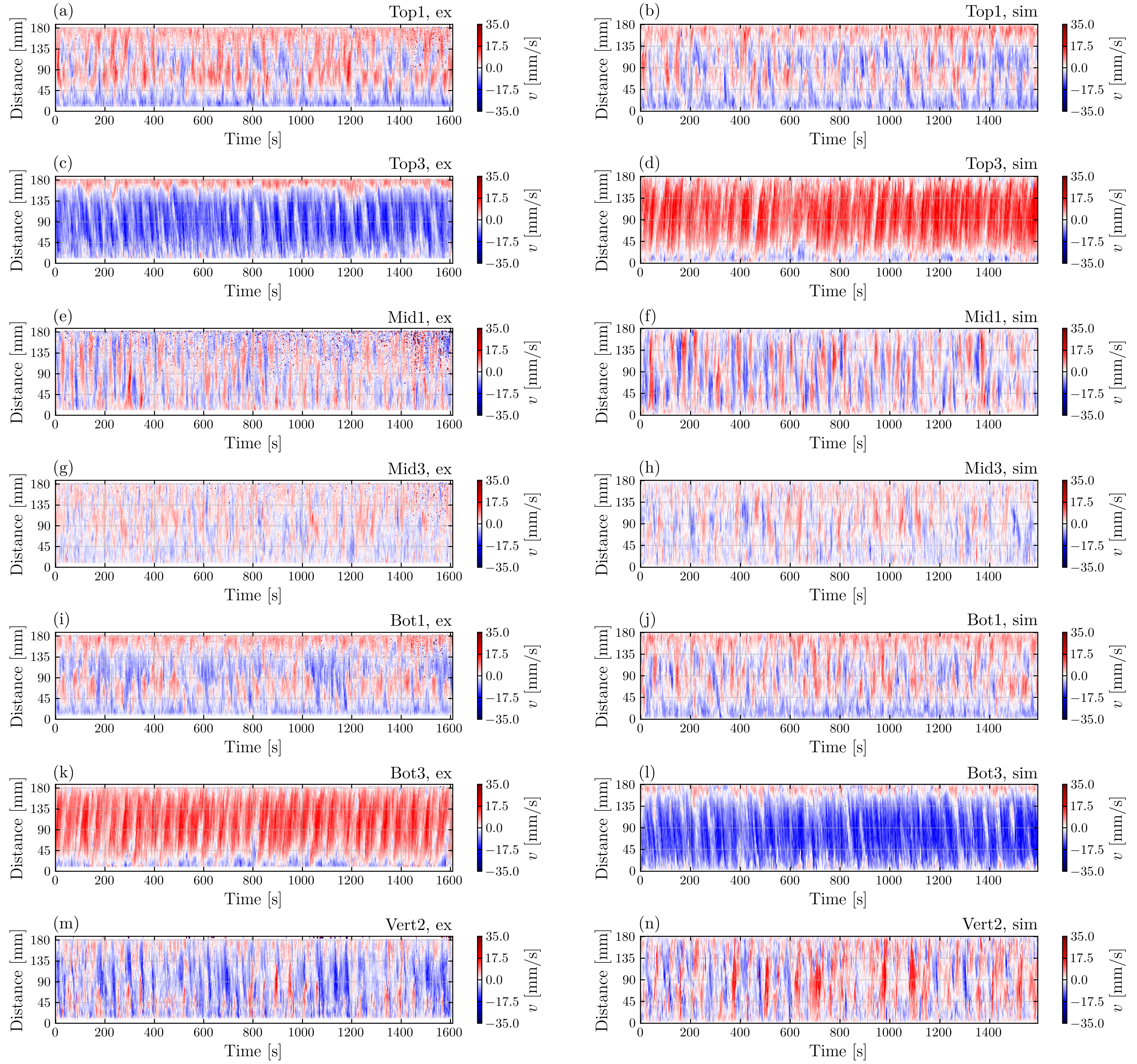}
	
	\caption{Same as Fig.\,\ref{resultsContour0A}, but for a coil current of \unit[21.2]{A}. \label{resultsContour21A}}
\end{figure*}

\begin{figure*}
	\centering
	\includegraphics[width=\textwidth]{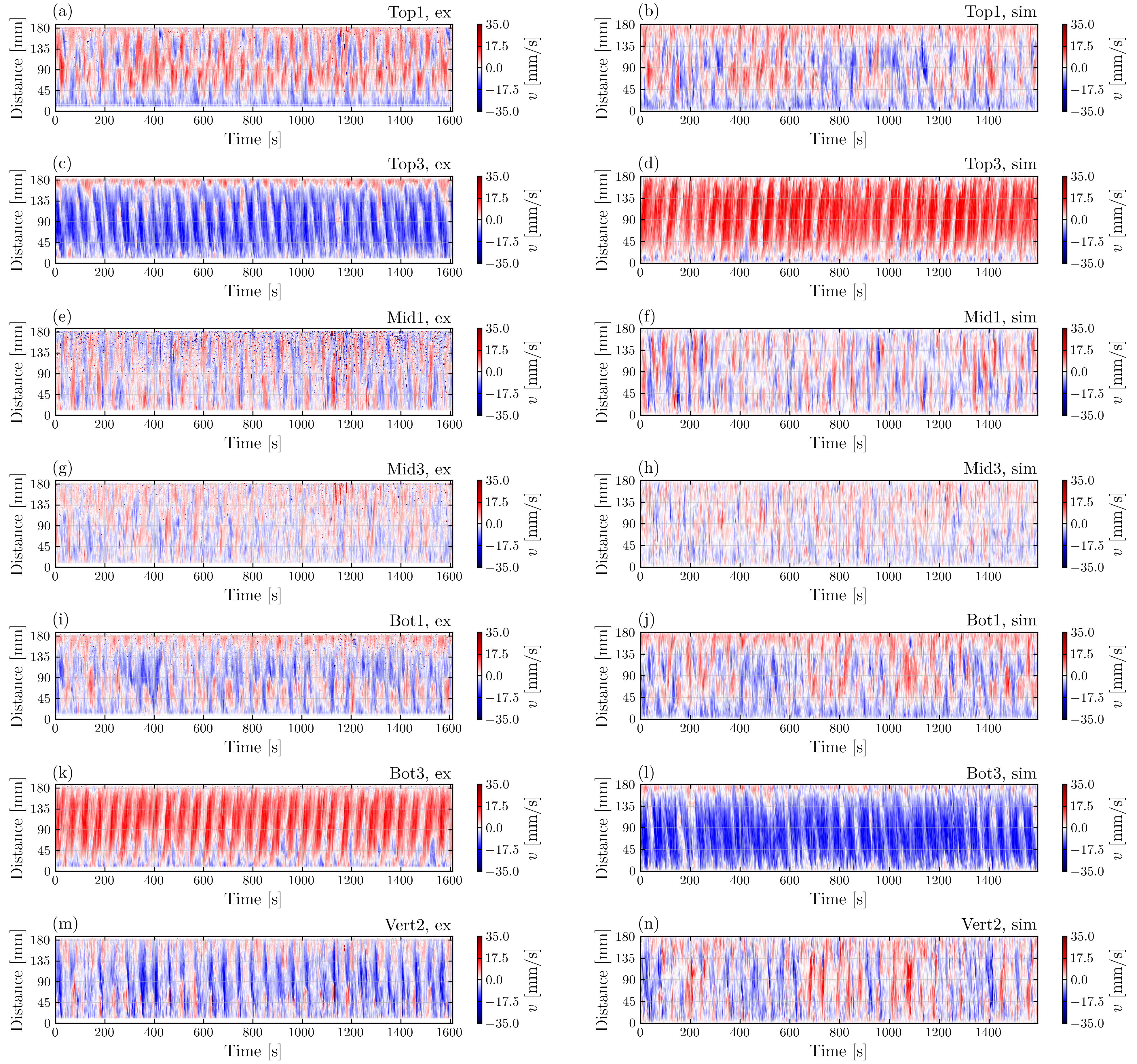}
	
	\caption{Same as Fig.\,\ref{resultsContour0A}, but for a coil current of \unit[27.6]{A}.\label{resultsContour27A}}
\end{figure*}

Although a wide variety of frequency, current and temperature parameters has been explored in the experiments, 
in the following we focus on a RB flow with a temperature 
difference $\Delta T= \unit[5]{K}$ which corresponds to a Rayleigh number of 
$Ra=1.03 \times 10^7$.

In all cases with forcing, the fundamental frequency of the coil current 
is fixed to \unit[25]{Hz}, while the amplitude is slowly modulated with
frequency $f_{mod}$. A setting of $f_{mod} = \unit[18]{mHz}$ has been chosen to be in the range of the natural sloshing frequencies of the LSC as mentioned before. The intention was, staying close to a possible 1:1 resonance, while being slightly off so a possible entrainment would be visible.\\
First, we present the UDV sensor data for four representative cases 
with increasing coil currents. 
This yields already a qualitative picture for the process under investigation.
A following spectral analysis and forcing-signal-correlations elucidate possible
causal relationships. Further insight is provided by evaluations of 
the simulated flow field, allowing for an attempt at a 
theory for the underlying mechanisms.

\subsection{UDV data for four different coil currents \label{resultsContourSection}}

Here we present the velocity profiles measured along
seven selected UDV sensor beams 
and compare them 
with the corresponding profiles of
{\it virtual sensors} as extracted from numerical simulations. 
Specifically, we discuss the results
for the four maximal coil current amplitudes of 
\unit[0]{A}, \unit[12.5]{A}, \unit[21.2]{A}, and \unit[27.6]{A}. The presented data are representative examples from, in some cases, much longer time series.

We start with the pure RB flow, i.e. with zero coil current. 
For seven selected sensors,
Fig.\,\ref{resultsContour0A} shows the contour plots 
(over time and UDV beam depth) of the
actually measured signal (left column) and of the corresponding 
``virtual sensors''  (right column). 
For the selected time frame of 1600\,s, all sensors show velocities with 
amplitudes of some
30\,mm/s, and short term oscillations with periods 
of appr. 50\,s.
The dominance of the flywheel-type LSC is clearly 
mirrored by the reciprocal flow direction (red versus blue) between sensors 
``Top1'' and ``Bot1'', as well as  between ``Top3'' and ``Bot3''. 
Whenever the flow 
is directed towards (blue) the sensor at the top,
the corresponding signal at the bottom sensor is red, 
indicating a motion away from the sensor. Appropriately, the ``Vert2''
sensor observes, for most of the time, a downward motion (red).
The signals from the 
``Mid'' sensors are noticeable weaker. What they actually
measure is not the 
flywheel itself, but basically its lateral deflection, which 
corresponds to the sloshing motion of the LSC
with its periodicity of appr. $50$\,s. On longer time scales not shown here,
we also observe an azimuthal drift of the main 
LSC direction, as well as sudden cessations with direction change. 
Qualitatively, the virtual sensors on 
the right hand side show a very similar behaviour, including the 
$\sim 50$\,s periodicity. Yet, 
in view of the random direction of the
LSC and its long-term drift, a perfect agreement of all details
cannot be expected. 

The next plot, Fig.\,\ref{resultsContour12k5A}, shows the corresponding
(real and virtual) sensor data for a coil current of 12.5\,A.
While neither the velocity amplitude nor the $\sim 50$\,s periodicity have
much changed from the 0\,A case, we observe now a clear regularization 
of the flow, with the LSC constantly oriented in $y$ direction, i.e., 
along the sensor path of ``Top3'' and ``Bot3''.
This effect has already been noted in \cite{Roehrborn2022a, Roehrborn2022b} 
and is confirmed in this dataset. The fact that the direction of the LSC
in the numerical simulation happens to be contrary to that in the experiment (red versus blue
in  ``Top3'') merely results from random initial symmetry breaking
and is of no consequence for the analysis.
Sloshing in $x$-direction is now most strongly expressed in the ``Mid1''
data, while the numerical data shows slightly more regularity than
the experimental one. As for the ``Vert2'' sensor, one can observe exchanging upward and downward streams resulting from the rotation/torsion of the LSC. 
Noteworthy is also the change in flow structure in the ``Top1'' and ``Bot1'' data. 
It switches from a mainly unidirectional flow (as in Fig.\,\ref{resultsContour0A}) 
to a four-section structure, which will become even more pronounced with 
increasing coil currents.

At \unit[21.2]{A} (Fig.\,\ref{resultsContour21A})
the regular patterns become obvious, 
particularly visible in the ``Top3'' and ``Bot3'' data. 
Despite being accordant to the forcing, as will be shown below,
they are unlikely to be just a measurement of the forcing,
as the particular sensors ``look'' perpendicular to the main forcing direction.
In particular, the unimodal flow direction is not what would be expected from
a purely forced flow, as in \cite{POF2020}.
Interestingly, also a somewhat higher frequency in ``Mid1'' emerges 
which will be discussed in more detail further below. 

Finally, the 27\,A case (Fig.\,\ref{resultsContour27A}) 
exhibits a very stable periodicity, 
pointing to a sort of asymptotic synchronization behaviour. 
The four-section structure  in ``Top3'' and ``Bot3'' is now very pronounced.

\subsection{Fourier spectra \label{sectionFFTs}}

Having observed an increasing flow regularity in the previous section, 
we asks now what frequencies are included and how they relate to the forcing.
For that purpose, we 
discuss the Fourier spectra of several locations in the cylinder. Specifically,
for ``Top3'', ``Bot3'', ``Mid1'' and ``Vert2''
we will focus on the central segments around 90\,mm, 
averaged over the depth between 85\,mm and 95\,mm.
For the ``Mid1''  sensor, we will additionally observe points
centered around 40\,mm, and 140\,mm, 
likewise averaged over the surrounding 10\,mm. Along the time axis, segments 
of ca. \unit[1600]{s} are selected identical to the ones displayed in section 
\ref{resultsContourSection}. The resultant time series are detrended and a 
von Hann window of the same length as the series is applied before the data are passed to NumPy's real-valued 
discrete Fourier Transform function ``rfft''.
The shown periodograms are to be interpreted as examples only. As 
the turbulent flow (of $Re=\bm u D/\nu\approx10^5$) has an inherently chaotic element to it, the results can
vary depending on the observed time frame (particularly for the low 
current values).

\begin{figure}
	\centering
	\includegraphics[width=0.48\textwidth]{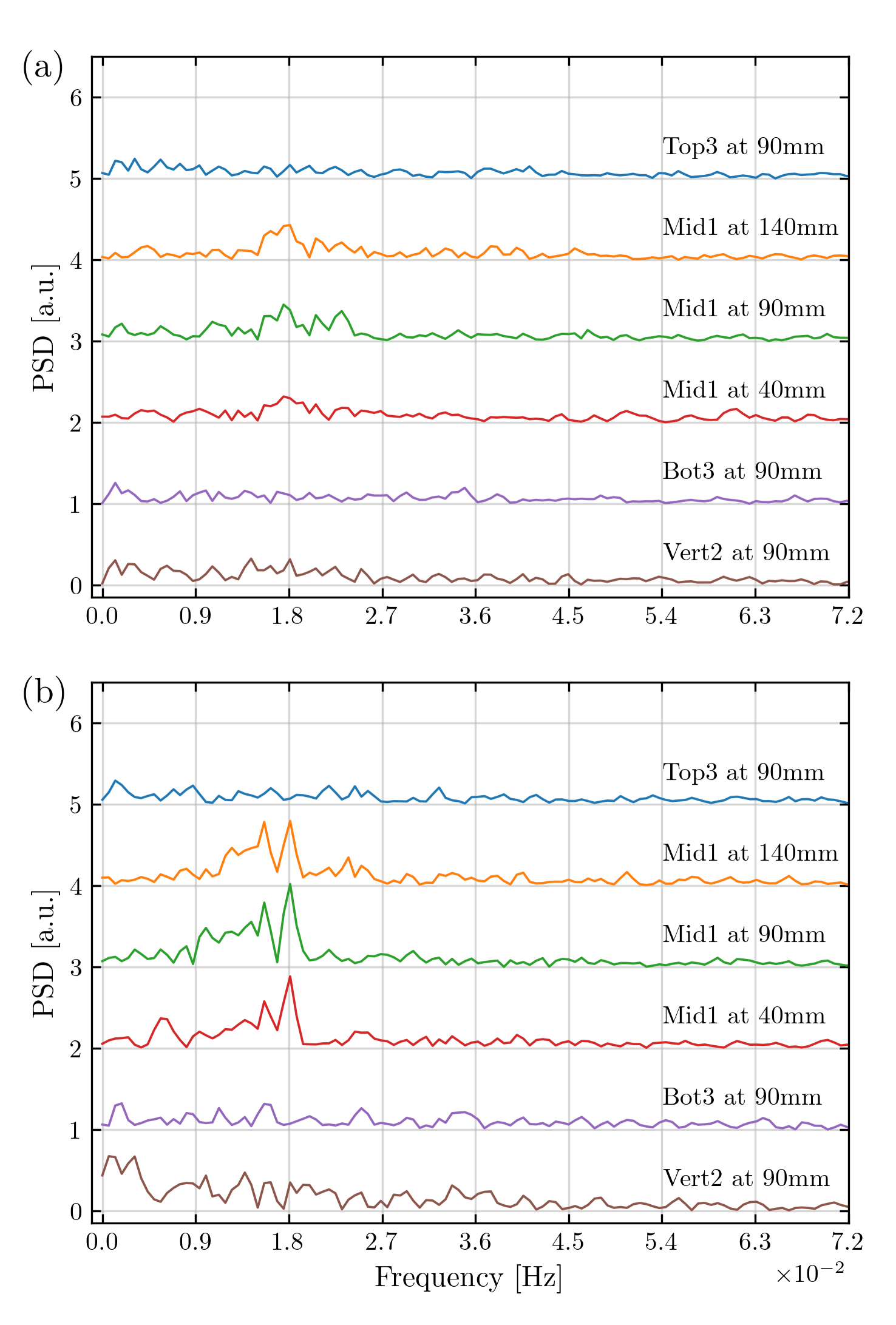}
	\caption{Power spectral density (partly shifted) 
	of six data sets from four sensors, extracted at various depths
	from Fig.\,\ref{resultsContour0A}, for a coil current of 0\,A.
	The upper panel corresponds to data from the real sensors,
	the lower panel to those of "virtual sensors" from simulation.
	\label{fft0A}}
\end{figure}

Let us begin, in Fig.\,\ref{fft0A}, with an analysis of 
the pure RB case ($0$\,A).
In view of long-term wanderings of the LSC in this case, 
the corresponding data in Fig.\,\ref{resultsContour0A} 
has been chosen from a longer time series 
such that the LSC is pointing roughly in the direction of the  ``Top3'' and 
``Bot3''sensors.
The FFTs of the center flow show a 
rather broad maximum around 18\,mHz (corresponding to
55.56\,s) which is the typical frequency of the sloshing/torsional
motion in our case. This is why it has been chosen as the forcing modulation 
period in this study. Such a peak actually can also be found 
in the angle data, a 
measure used in previous investigations \cite{LSC-Till-2019}.

\begin{figure}
	\centering
	\includegraphics[width=0.48\textwidth]{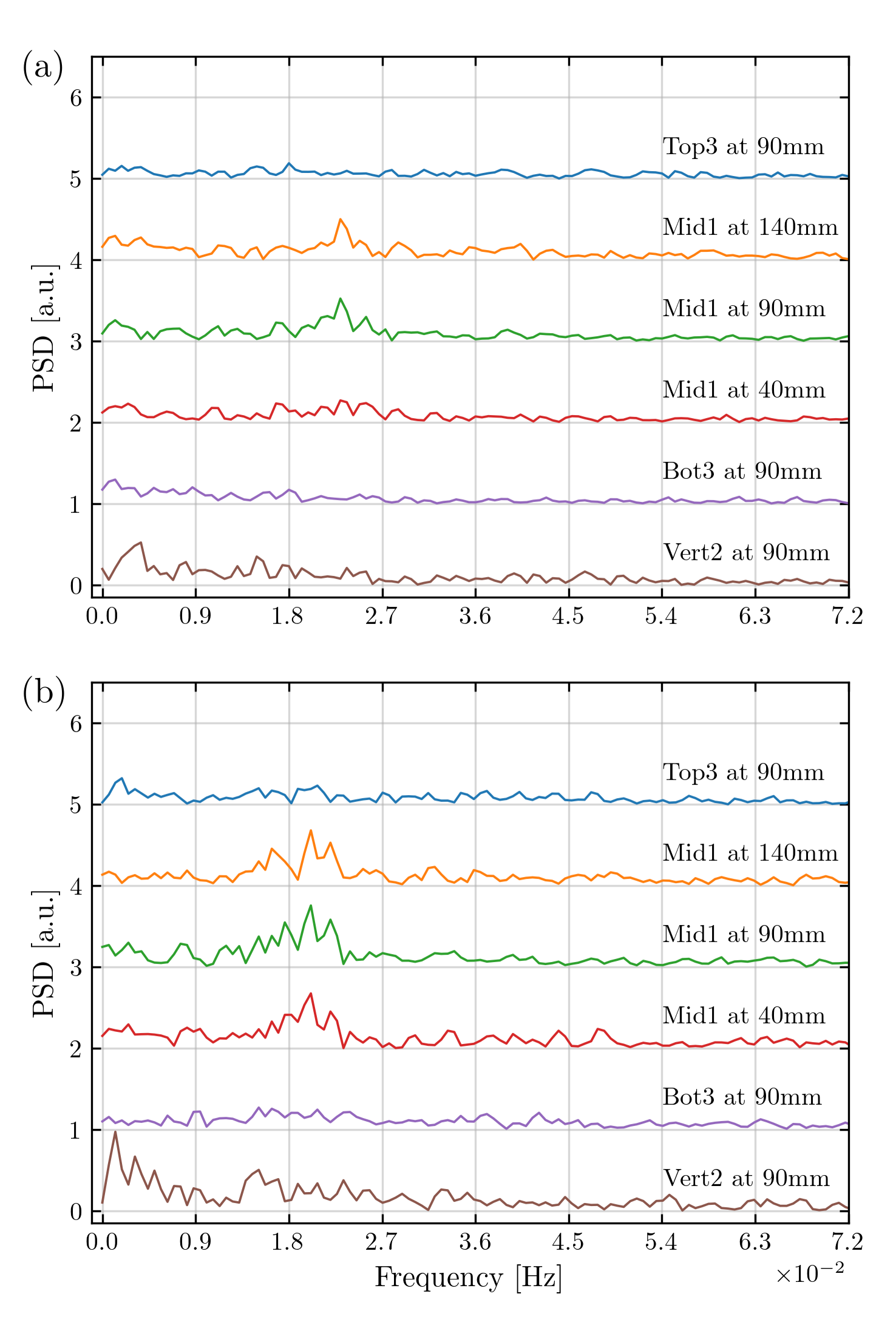}
	\caption{Same as Fig.\,\ref{fft0A}, but for 12.5\,A.
	\label{fft12k5A}}
\end{figure}

At \unit[12.5]{A} the peaks of the ``Mid1'' sensor
start to shift towards higher frequencies, as can be seen in 
Fig.\,\ref{fft12k5A}. The other measurement points still seem to 
be much too irregular in behaviour for the FFT to pick up anything 
worth mentioning.

\begin{figure}
	\centering
	\includegraphics[width=0.48\textwidth]{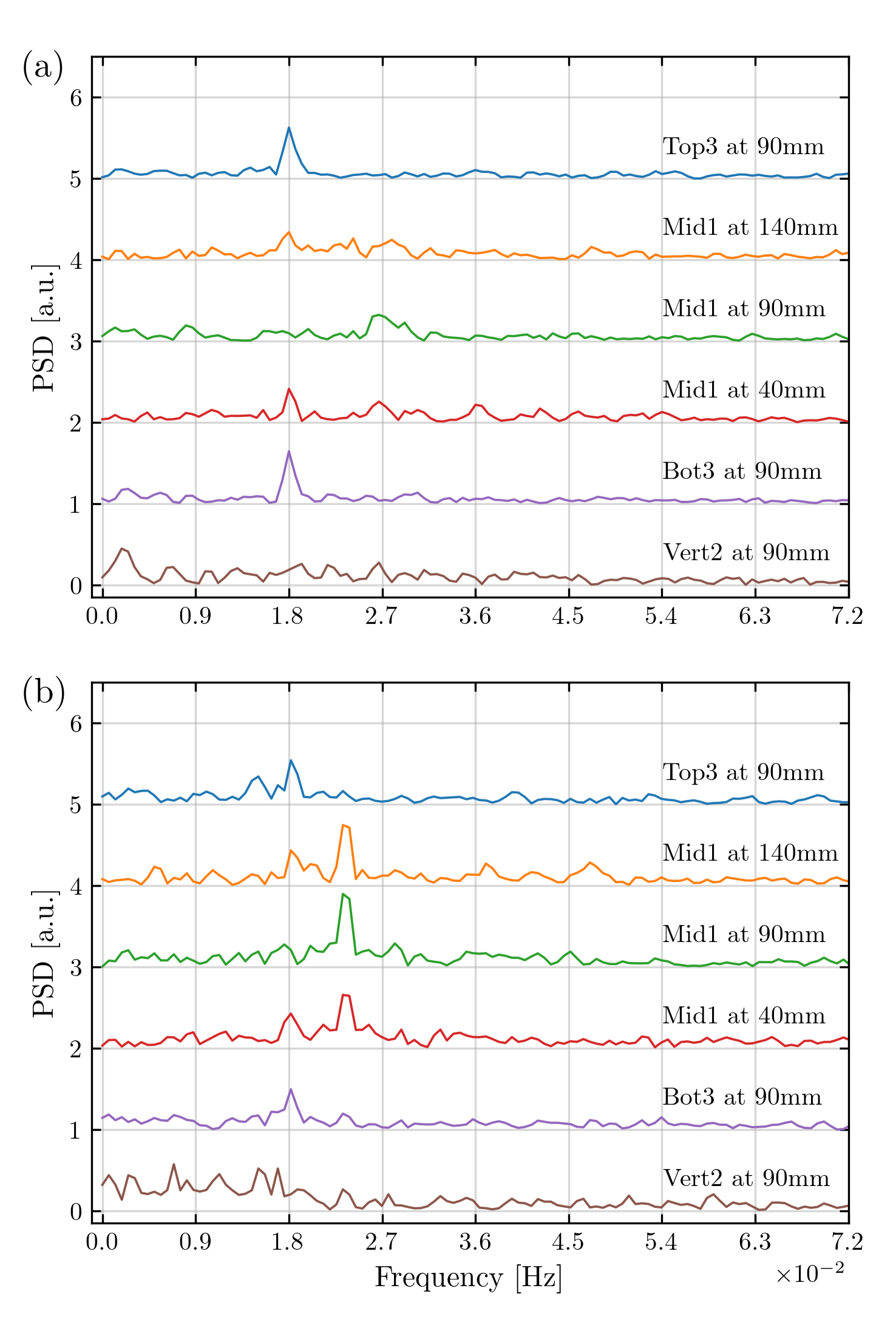}
	\caption{Same as Fig.\,\ref{fft0A}, but for 21.2\,A.
	\label{fft21k5A}}
\end{figure}

Increasing the coil current further to \unit[21.2]{A},
Fig.\,\ref{fft21k5A}
displays now prominent peaks at the forcing frequency of 
\unit[18]{mHz}, both for experiment and simulation. The points roughly 
one fifth from the wall in ``Mid1'' also see this frequency in addition to 
the faster periods. Those faster periods also shift a little further up the scale, 
which seems to be consistent behaviour.

\begin{figure}
	\centering
	\includegraphics[width=0.48\textwidth]{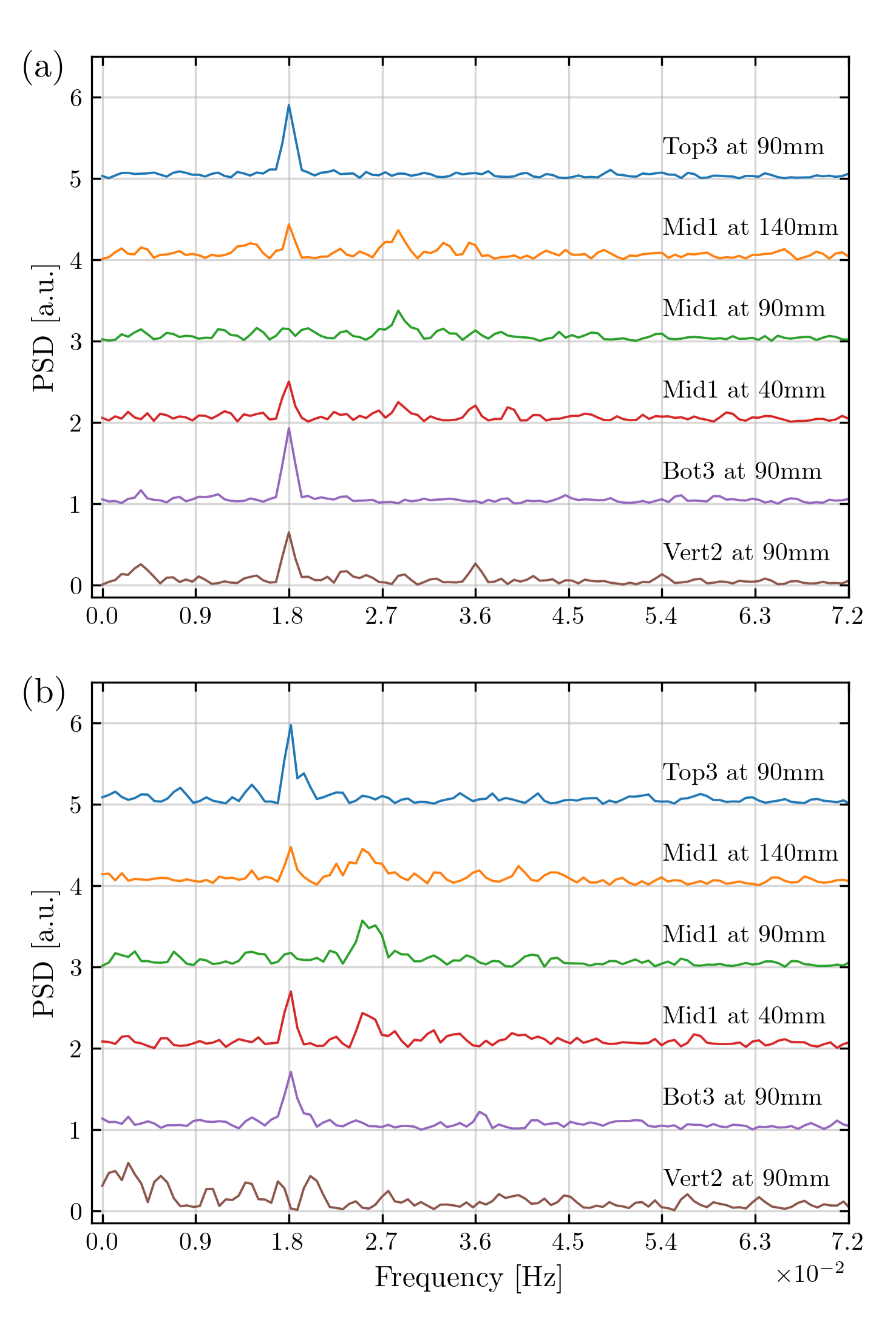}
	\caption{Same as Fig.\,\ref{fft0A}, but for 27.6\,A.
		\label{fft27A}}
\end{figure}

Further increasing the coil current to 27.6\,A 
shows the same pattern with an even more 
expressed forcing 
period in the ``Top3'' and ``Bot3'' data (Fig.\,\ref{fft27A}). 
At this stage it is even  
expressed in the ``Vert2'' experimental data.

\begin{figure}
	\centering
	\includegraphics[width=0.48\textwidth]{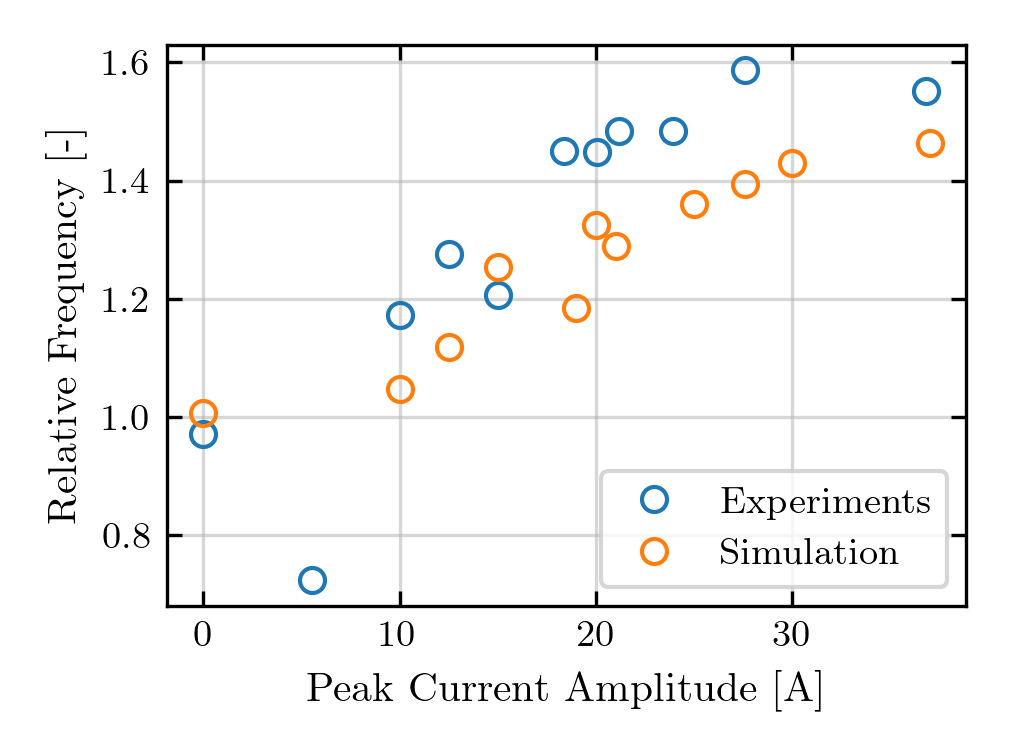}
	\caption{Analysis of the strongest frequencies in the Mid1 sensor at
		 \unit[90]{mm} for experiments and numerics, relative to the forcing
		  modulation frequency of \unit[18]{mHz}. The data points at
		   \unit[37]{A} are questionable, as the peaks are not at all
		    pronounced.\label{midfrequencies}}
\end{figure}

We conclude that the increasing Lorentz-forces clearly leave their mark 
in the FFT spectra as well. A remaining question is, how the fluctuation at the cylinder centre evolves.
As a first glance, Fig.\,\ref{midfrequencies} depicts the strongest frequencies for the 
\unit[90]{mm} data divided by the modulation frequency of \unit[18]{mHz}. 
It is not a perfect measure and subject to statistical fluctuations, 
but suggests a continuous increase.
And while the \unit[37]{A} data points are not quite reliable due to small peaks, further simulations 
suggest a saturation beyond that current. At any rate, 
the observed periods are somewhat 
unstable and thus the frequency peaks are not sharply expressed.

\subsection{Transition to synchronization \label{sectionSynchronization}}

With a first characterization at different forcings at hand, 
we now discuss the synchronization in a more quantitative manner. 
For that purpose, we assess the correlation between the
force signal $M(t)$ according to Eq.\,\ref{modulation} and the 
velocity around the centre of ``Bot3''. The latter data is chosen because the 
measured LSC velocities are perpendicular to the forcing 
and the sensor is delivering a low noise signal throughout all measurement runs. 
It also shows the strongest forcing response in the contours \ref{resultsContour0A}-\ref{resultsContour27A}.
The employed metric is Pearson's empirical correlation coefficient:
\begin{equation}
	r_k(x, y) := \frac{
		\sum_{i=1}^n(x_{i}-\overline x)(y_{ik}-\overline y_k)
	}{
		\sqrt{
			\sum_{i=1}^n(x_{i}-\overline x)^2
			\sum_{i=1}^n(y_{ik}-\overline y_k)^2
		}
	}
\end{equation}
where $x$ is a 27 periods long segment of the forcing envelope curve, 
y is the formerly mentioned velocity data, from which a 
29-period length is selected, 
and k denotes the lag between the correlation signals.
During correlation, the signal y is shifted across signal x and only the part of x is used that has 
a corresponding value in the y data. In order for this to be executable, both signals exhibit an identical temporal sampling.
 
As the forcing signal is 27 periods ($\equiv$\unit[1500]{s}) long, only a 
long-term coherent effect will yield a high correlation coefficient 
and thus hint at synchronization. If the synchronicity is weak 
and the periodic velocity's phase drifts with respect to the forcing 
or looses its periodic structure, the correlation coefficients will be small.

As two periods of lagging were chosen, when plotted over the lag distance 
the resultant coefficients resemble more ore less two periods of a sinusoid 
for all datasets (including plain RB). Thus, the maximum correlation and 
average phase difference can be estimated. Non-linear fitting of a 
sinusoid has been used to estimate the phase lag.

\begin{figure}
	\centering
	\includegraphics[width=0.48\textwidth]{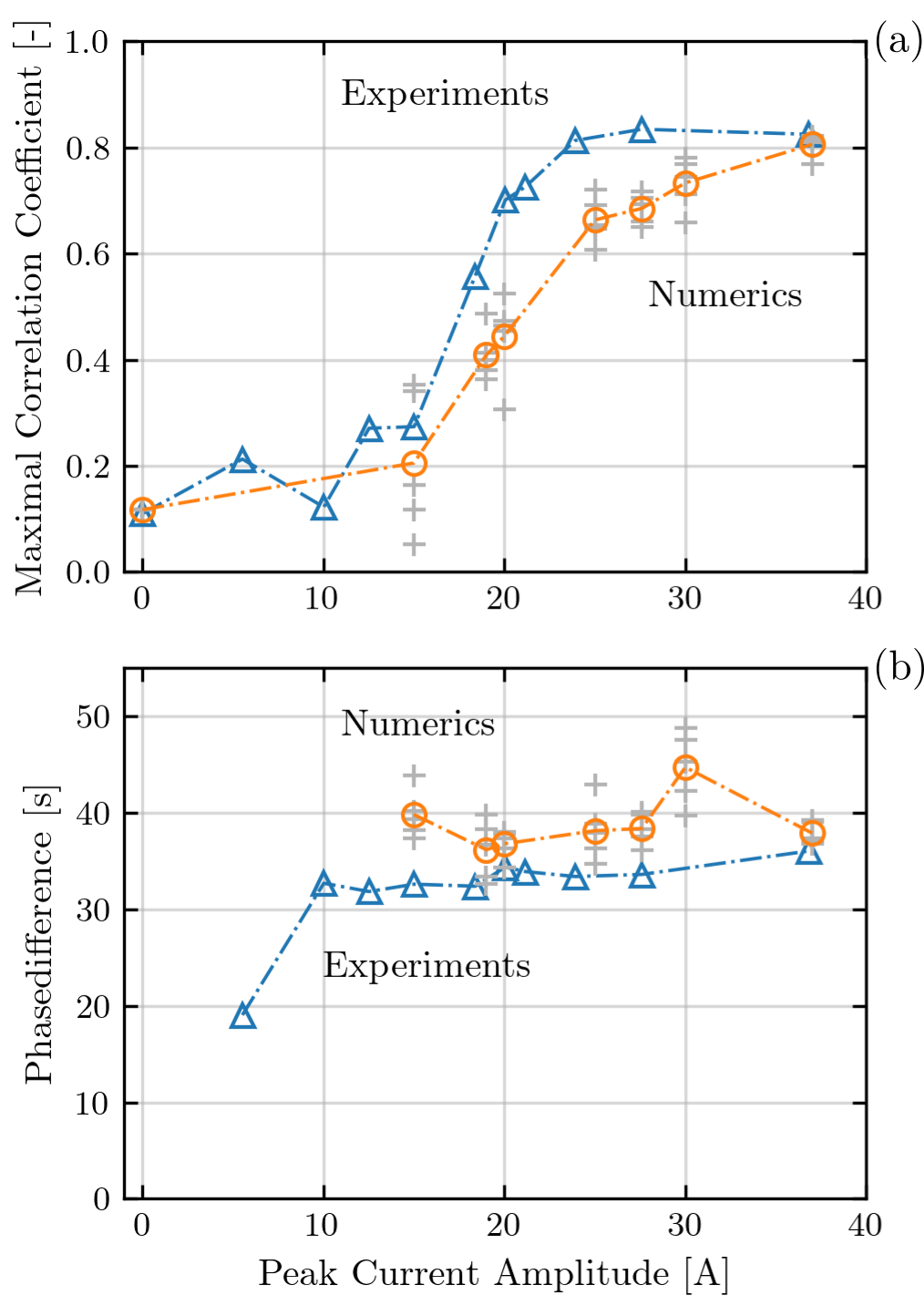}
	\caption{Onset of synchronization with increasing current.
		(a) Maximum of the empirical correlation coefficient of 
		sensor ``Bot3'' with the function $M(t)$ according to Eq.\,(5).
		(b) Phase shift between the two signals where the maximum 
		of the correlation occurs.
		In both panels, the blue triangles represent experimental results, while the grey crosses represent the numerical results for multiple intervals on the same dataset. The orange circles depict the average over the crosses for each current amplitude.
		\label{correlationBoundary}}
\end{figure}

In Fig.\,\ref{correlationBoundary} those maximal coefficients and 
corresponding phase lags are plotted against the coil current. 
The blue triangles are taken from measurements, while the grey crosses are
values from the numerical data. The orange circles represent the average of
the grey crosses for one current value.

Evidently, at low currents there is almost no correlation between the forcing 
and the flow (for the plain RB case, a proxy sine function with 
\unit[18]{mHz} was used). Then,
the correlation rises steeply between 10\,A and 25\,A, after which it reaches 
a sort of plateau.
As long-term coherence is required for such high values
to occur, we take this as strong evidence for a synchronization effect at play.

Nevertheless, the phase difference is quite large with more than half a period lying between the maximum of the forcing and the maximum of the flow velocity. This might hint at the synchronization mechanism working in a part of the cylinder which is away from the centre.

While the phase in Fig.\,\ref{correlationBoundary}(b) appears to be 
very steady even starting at \unit[10]{A}, there is, obviously
no reasonable phase relation to be defined for lower currents.
Since the ``coherence length'' is rather short, 
the maximal correlation coefficient is 
strongly dependent on the chosen time frame. 
A first attempt on characterizing the underlying statistics has been made using a sort of bootstrapping on the numerical data. 
There, longer datasets were available 
and more than one interval of \unit[1500]{s} could be evaluated. 
The result are the grey crosses in Fig.\,\ref{correlationBoundary}. 

For high currents, the crosses have a consistently small spread, 
which gives some validity to the former interpretation of the data.
Yet, the correlation does not reach 
exactly one, which is due to noise from turbulence always contained in the data. 

Even though a synchronization effect is readily apparent, 
the actual process facilitating it has to be further illuminated.

\subsection{Flow geometry and helicity synchronization \label{sectionFlowGeometryHelicity}}

In this section we will have a closer look on the actual mechanism of the synchronisation. 
For that purpose, we mainly rely on the numerical simulations which allow to analyze the
flow in much more detail than it would be possible from the limited number of sensors
used in the experiment. This applies, in particular, to the helicity of the flow
which can only be quantified by numerical data. 
However, the good correspondence between numerics 
and experiments presented so far makes us confident about the
physical relevance of the numerical simulations.

\begin{figure*}
	\centering
	\includegraphics[width=0.8\textwidth]{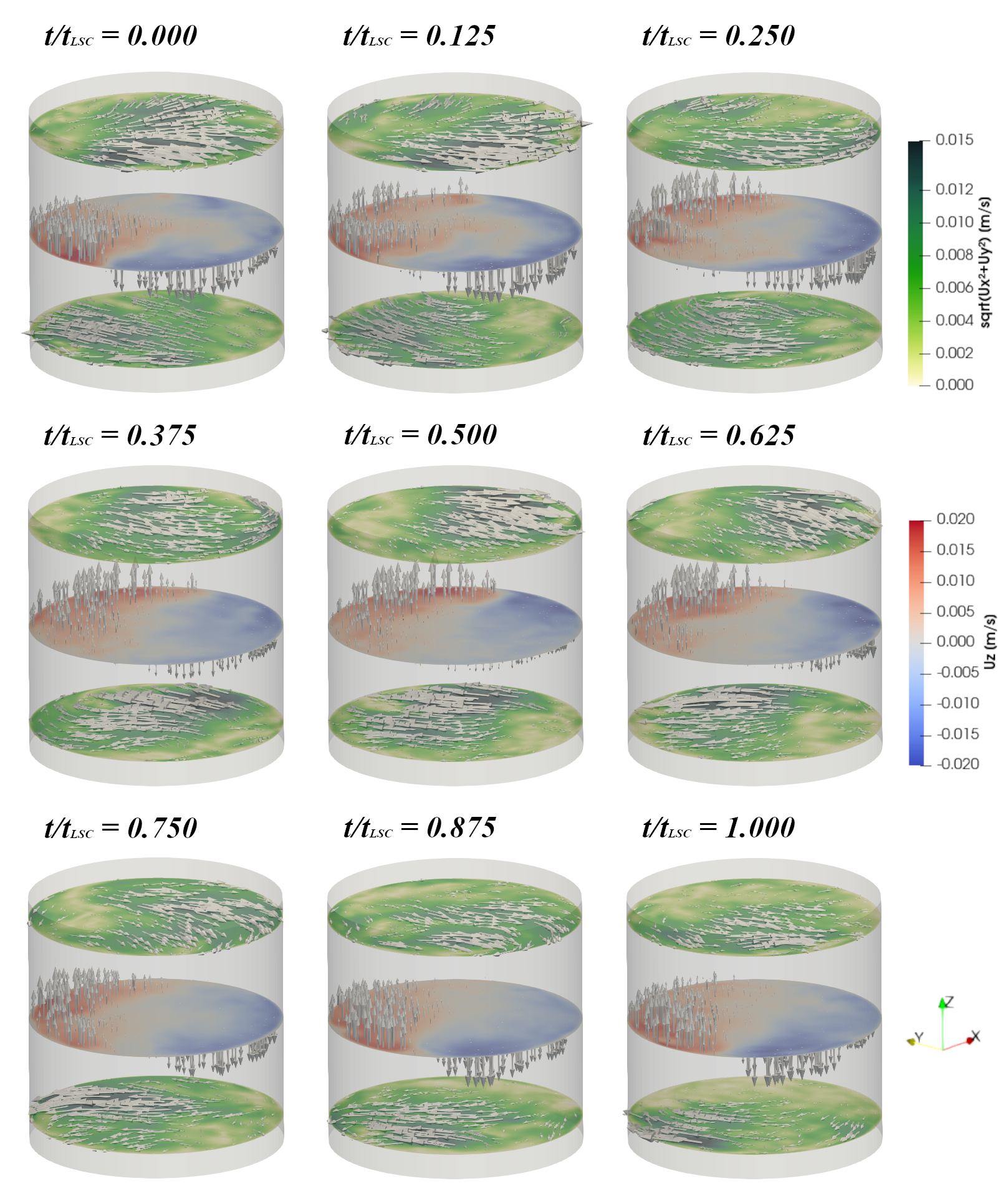}
	\caption{Illustration of the vector flow field for a coil current of 
		\unit[0]{A} (pure RB convection).
		The 9 panels cover a typical sloshing period of 55 
		seconds, during which the LSC undergoes a ``round trip''.
		The upper and lower planes show the horizontal component
		of the velocity (gray arrows), including its intensity (color).
		The plane at mid-height shows the corresponding vertical 
		velocity component.	\label{illustration0A}}
\end{figure*}

\begin{figure*}
	\centering
	\includegraphics[width=0.8\textwidth]{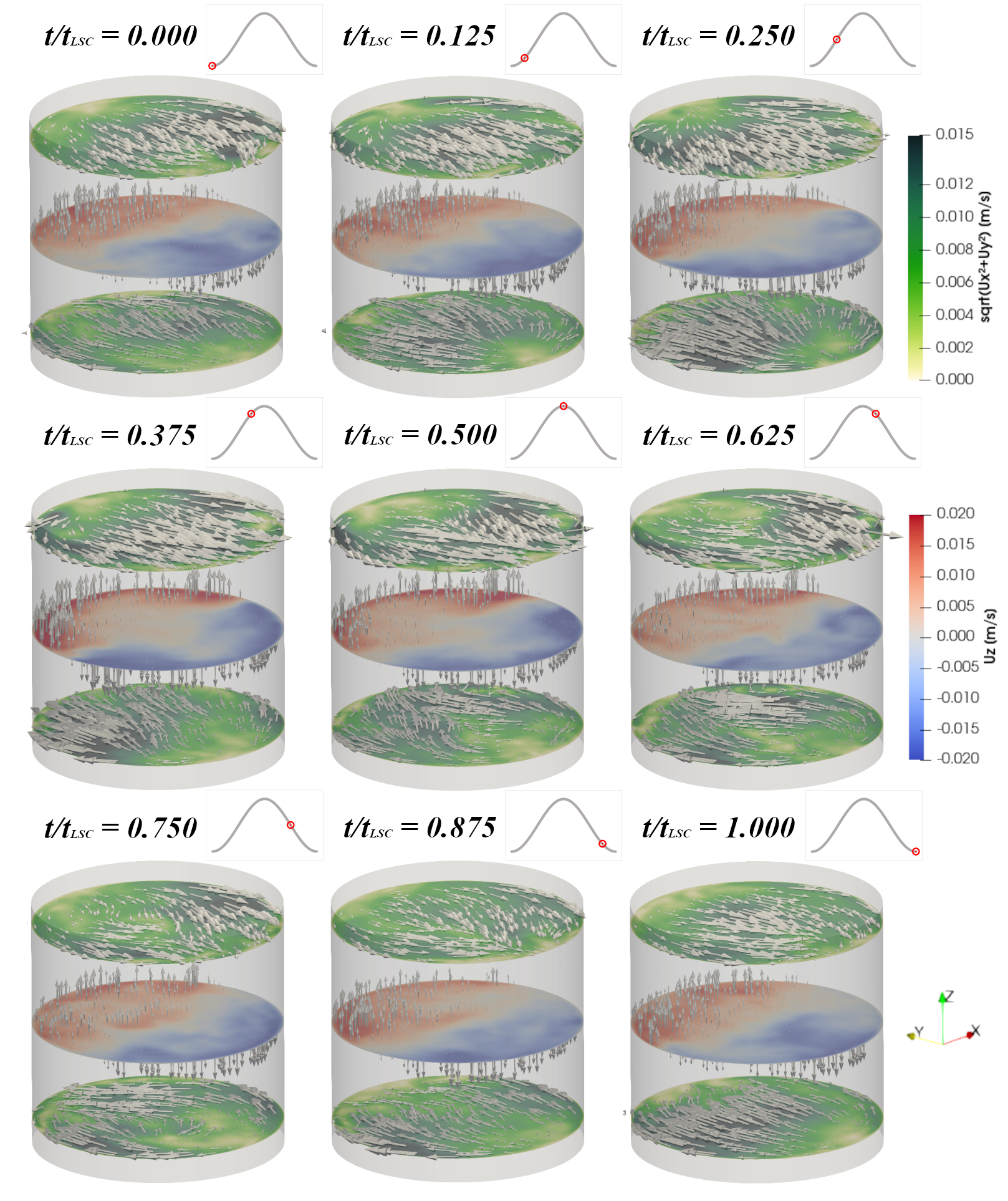}
	\caption{Same as Fig. \ref{illustration0A}, but for a coil 
		current of \unit[30]{A}.\label{illustration27k6A}}
\end{figure*}

As previously stated, the flow geometry changes from a plain
RB case to a forced case. 
In Figs.\,\ref{illustration0A} and \ref{illustration27k6A}, 
the flow is illustrated in 9 panels distributed 
over one period of \unit[55]{s}. 
For the pure RB case, in Fig.\,\ref{illustration0A} the panels 
show the typical sloshing and
torsional motion associated with the movement of the LSC in the volume. 
The forced case with $30$\,A, shown in Fig.\,\ref{illustration27k6A},
reveals a more complex behaviour. 
On the mid-height plane, a back and forth motion is visible but has additional components superimposed. 
On the top and bottom planes, the flow seems to be parted in two, 
along positive and negative $x$, whenever the forcing starts to act.
On closer inspection it appears that the LSC is strong on one side
and then switches sides without visibly crossing the center over the course of one period.
The large scale structure of the LSC seems to become more complex with 
those branching flows.

There are several ways to interpret this behaviour. 
Looking at how the alternation of sides in plume formation is regularized, 
an interaction of the Lorentz-forces with the internals of the RB process is suggested. 
Further investigation is required to exclude this possibility.

\begin{figure*}
	\includegraphics[width=0.5\textwidth]{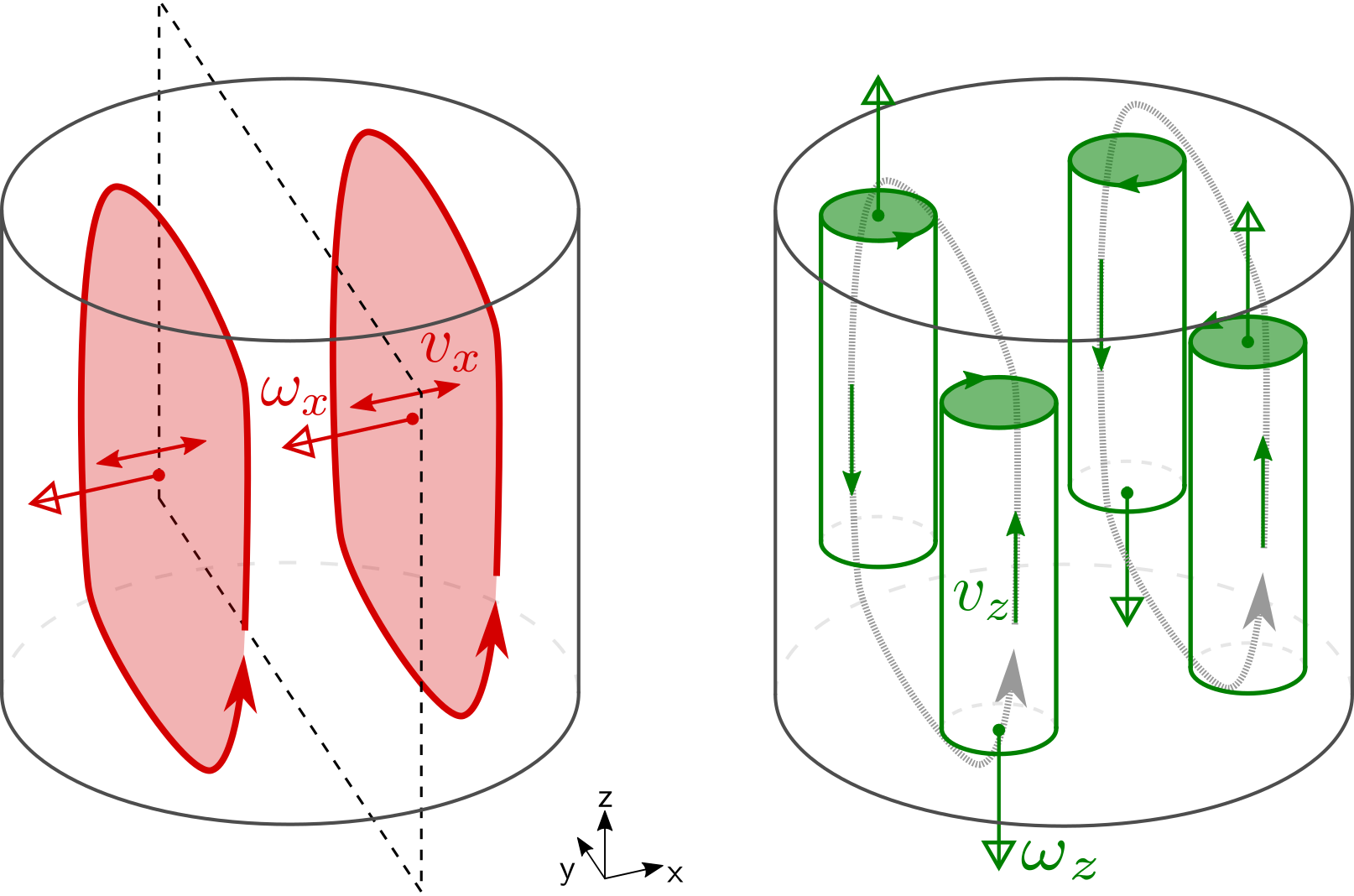}
	\caption{Illustration of the main contributions to the helicity, 
	$h_x=u_x \omega_x$ (red) and $h_z=u_z \omega_z$ (green). 
	\label{rolle}}
\end{figure*}

Nevertheless, turning to the helicity we can find another mechanism. In order to understand this concept, we have to make a few distinctions beforehand which are
illustrated in Fig.\,\ref{rolle}.
In addition to the full helicity\cite{Moffatt2018} 
\begin{equation}
    H = \int_V \bm u \cdot \omega dV, \quad \mathrm{with}\ \omega = \nabla \times \bm u
\end{equation}
integrated over the entire volume $V$, we consider
also the partial helicities $H^{-}$, integrated over the restricted volume with $x<0$,
and $H^{+}$, integrated over $x>0$. Furthermore, we distinguish between the two
helicity density contributions $h_x=u_x \omega_x$ (red) and $h_z=u_z \omega_z$ (green),
whose volume integrals are denoted by $H_x$ and $H_z$, respectively.
Roughly speaking, the first one, $h_x$, represents the projection of the 
vorticity $\omega_x$ of the LSC 
on its sidewise motion $u_x$
(presupposing that the LSC is mainly directed in $y$-direction, which is 
safely guaranteed 
only in the synchronized regime).
The second one, $h_z$, represents the projection of the
vorticity $\omega_z$ of the four rolls (arising from the m=2 forcing \cite{POF2020})
on the vertical component $u_z$ of the LSC which 
penetrates them. The third contribution, 
$h_y=u_y \omega_y$ has not such a clear interpretation and is, therefore,  
skipped in the following.

\begin{figure*}
	\centering
	\includegraphics[width=\textwidth]{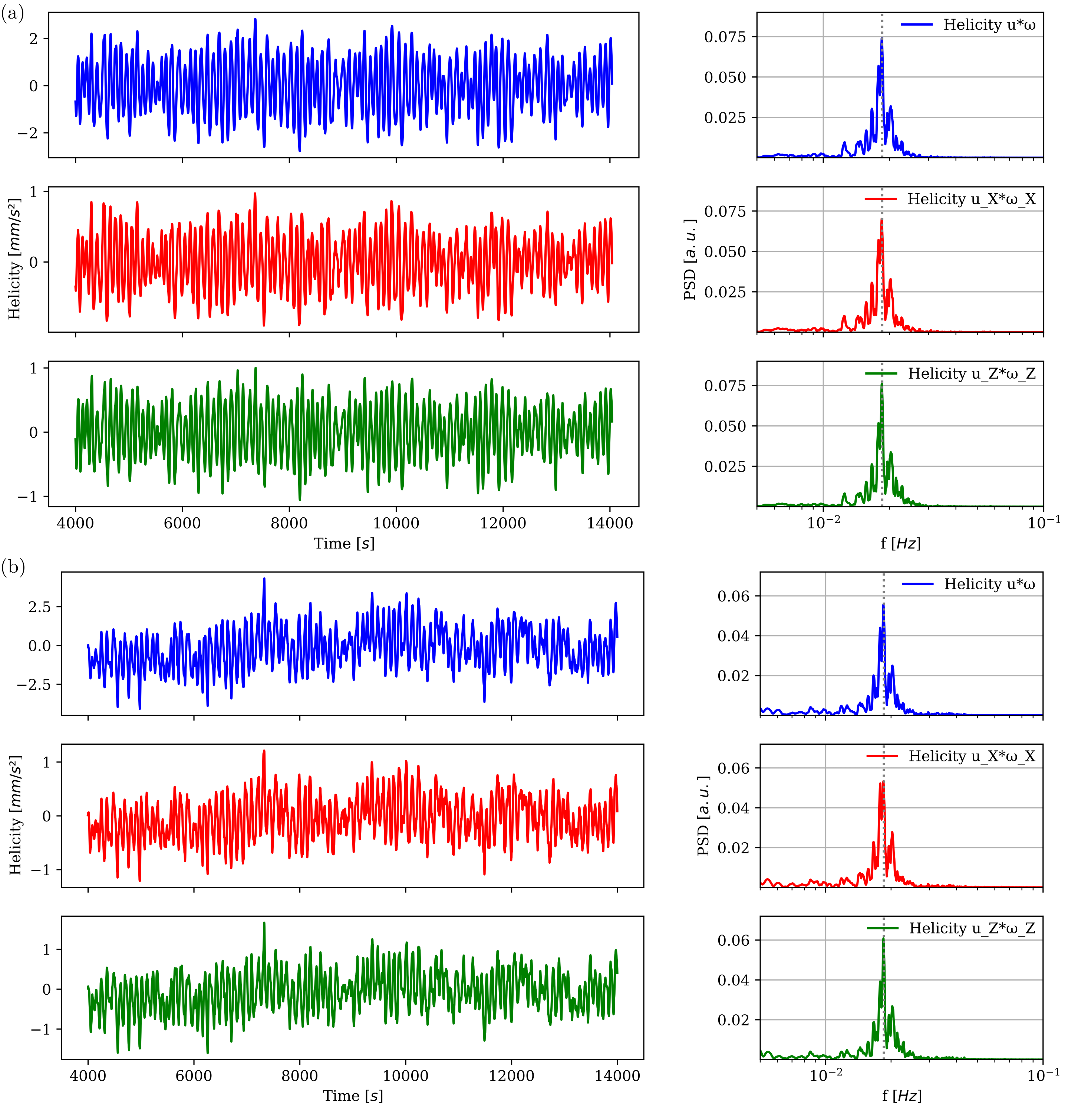}
	\caption{Time series (left) and FFTs (right) of the helicities $H$ (here normalized by the volume) 
		in the plain RB flow. Pane (a) contains helicities calculated 
		over the full volume, while in pane (b) only one half cylinder
		 (negative x-values) is considered. The rows in each pane relate 
		 the full helicity $H$ (blue) and the partial contribution $H_x$ 
		 (red) and $H_z$ (green). The other half of the cylinder (positive
		  x-values) yields comparable results and is therefore not shown. 
		  \label{heli0A}}
\end{figure*}

\begin{figure*}
	\includegraphics[width=\textwidth]{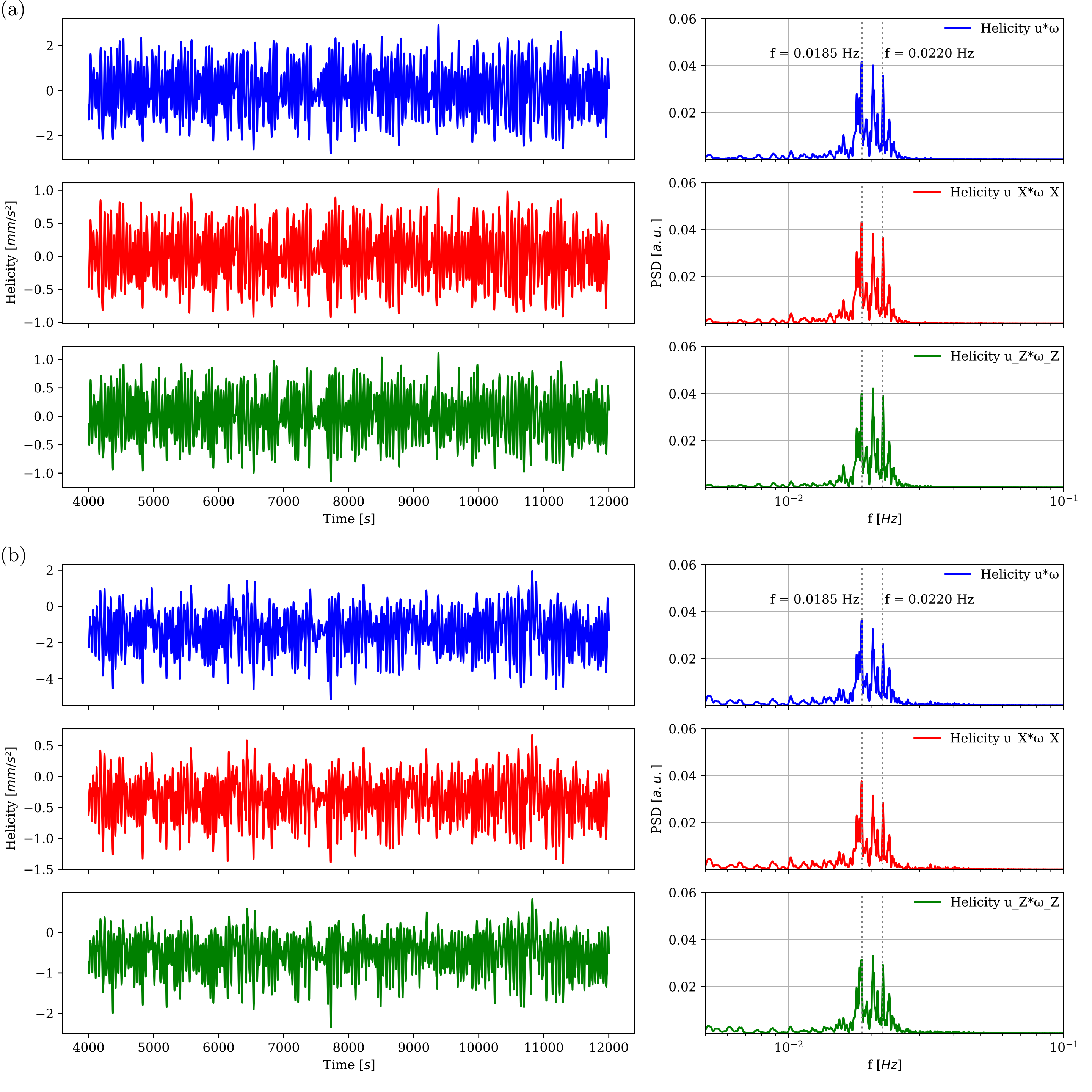}
	\caption{Same as Fig.\,\ref{heli0A}, but for a coil current of \unit[12.5]{A}. \label{heli12A}}
\end{figure*}

\begin{figure*}
	\includegraphics[width=\textwidth]{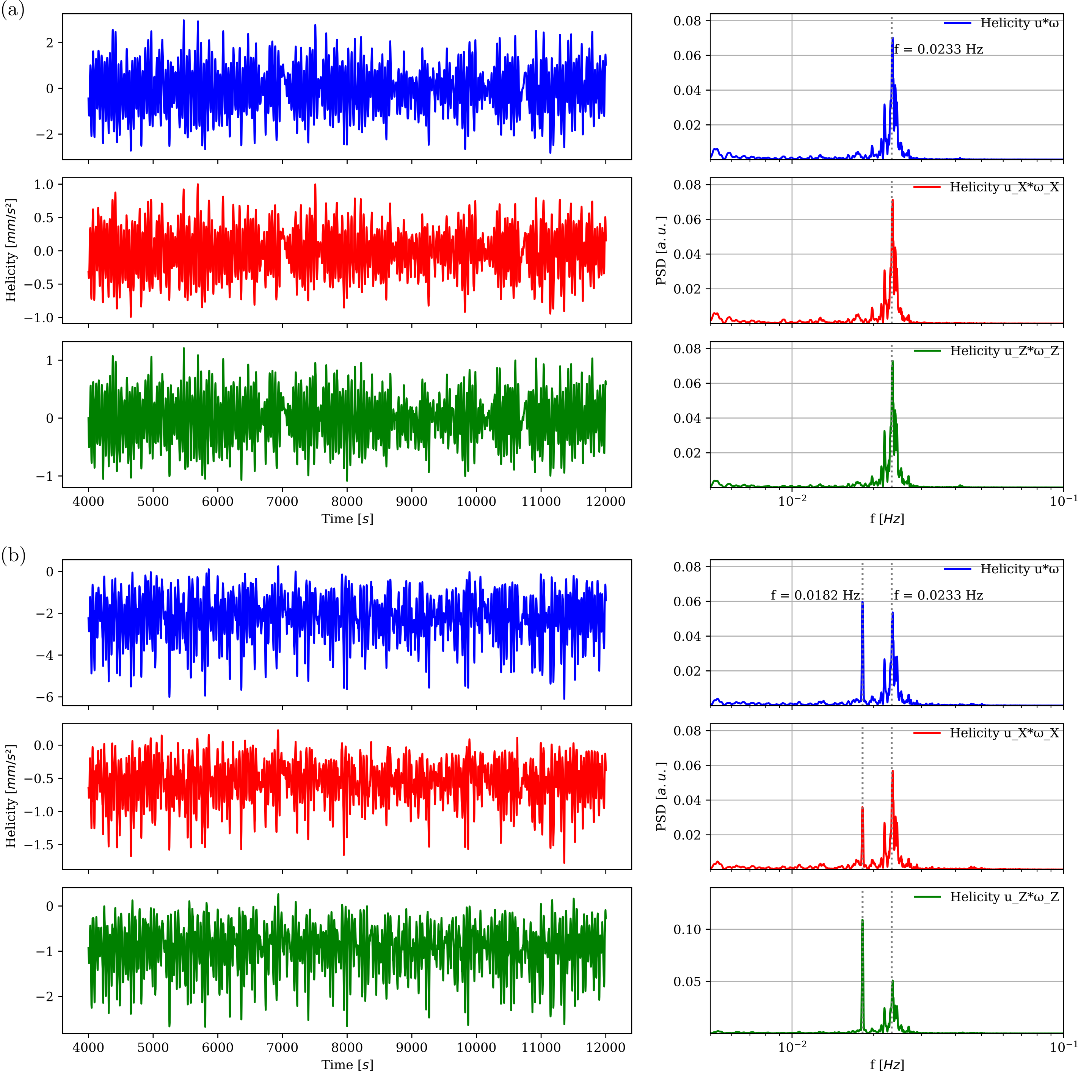}
	\caption{Same as Fig.\,\ref{heli0A}, but for a coil current of \unit[21.5]{A}. \label{heli21A}}
\end{figure*}

\begin{figure*}
	\includegraphics[width=\textwidth]{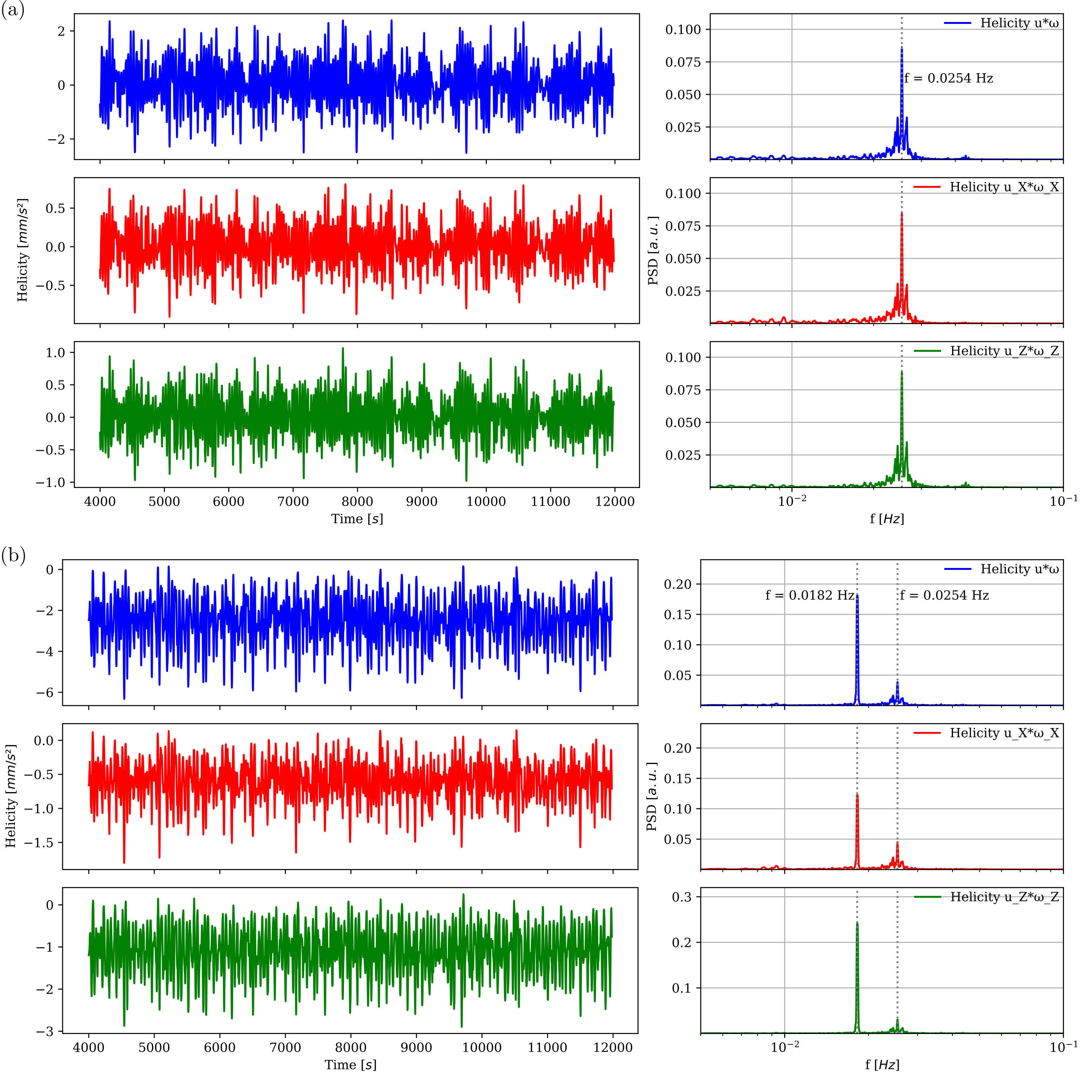}
	\caption{Same as Fig.\,\ref{heli0A}, but for a coil current of \unit[27.6]{A}. \label{heli27A}}
\end{figure*}

We start again, in Fig.\,\ref{heli0A}, with the pure RB case.
Its left column exhibits time series of various helicity
components, the right column shows the corresponding
FFTs. The three rows of panel (a) show helicities averaged over the entire volume, 
while the rows of panel (b) show the 
corresponding partial helicity $H^{-}$
for the restricted volume with 
$x<0$ (we skip $H^{+}$ for $x>0$ as its graph looks
similar to $H^{-}$, except with opposite sign).
What we observe here in all FFTs is a rather broadband distribution,
mirroring the sidewise motion of the LSC with 
its main, but not very sharp sloshing frequency.

For the next case of 12\,A (Fig.\,\ref{heli12A}), things are starting to change.
All FFTs now show peaks at the
synchronizing frequency of 18\,mHz and 
also previously mentioned higher frequencies 
(see section \ref{sectionFFTs}). 

Going over to the case with 21.5\,A as shown in Fig.\,\ref{heli21A}, 
we observe even more drastic changes.
All FFTs now exhibit extremely sharp peaks but with a decisive difference between the 
full-volume helicity $H$ and the half-volume helicity $H^{-}$. In the former, 
only a higher frequency at \unit[23]{mHz} is apparent, while in the latter, a strong and narrow peak remains at the forcing frequency of 
\unit[18]{mHz}.

At \unit[27.6]{A}, which corresponds the to the 
``fully synchronized state'', 
the distinction becomes more or less complete. 
In difference to the full-volume version, the half-volume helicity $H^{-}$ 
of Fig.\,\ref{heli27A}(b) are now dominated by the frequency of the forcing 
modulation, with only a minor component at the higher frequency.

\begin{figure*}
	\includegraphics[width=0.75\textwidth]{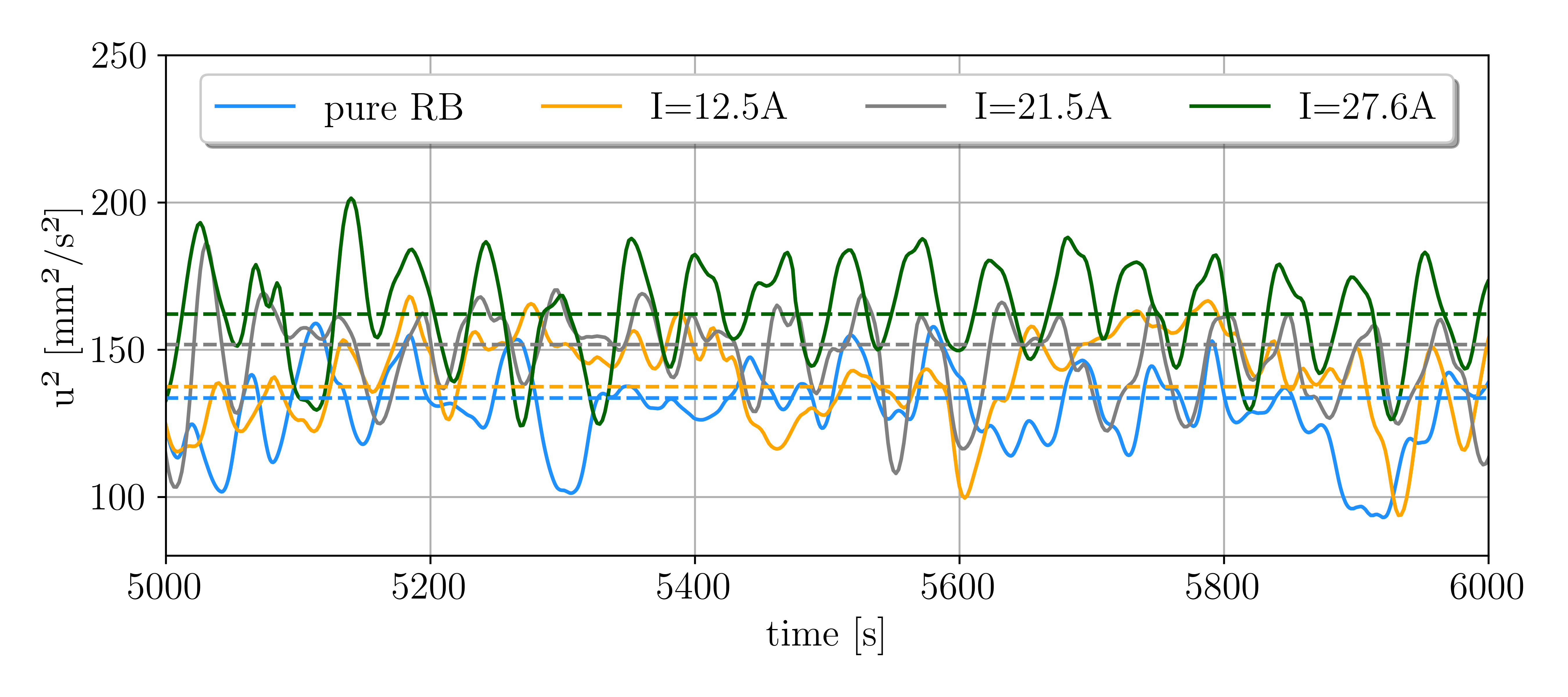}
	\caption{Mean kinetic energy of the flow for the four cases considered in this
	paper.  
	\label{uquadrat}}
\end{figure*}

From this evidence we conclude that helicity synchronization indeed occurs. 
The two half-volume 
helicities $H^{-}$ and $H^{+}$ are synchronized by the 
external $m=2$ forcing. $H_x$ and $H_z$ contribute comparable 
shares and take control of the cycle. It is not only the (unsurprising)
oscillation of $\omega_z$ of the four rolls directly driven by the 
$m=2$ forcing which in combination with the reciprocal sign change of $u_z$ translates to a non-zero signal. But also the
oscillation of $u_x$ leads to a more or less
symmetric sloshing of the LSC in the two x half-spaces. 
This explains also the clear peak at the forcing frequency 
as seen at the ``Top3'' and ``Bot3'' sensors in Fig.\,\ref{fft27A}, evidencing that 
the split LSC 
``passes by'' at $x=0$ with that frequency.
Evidently, with decreasing forcing this dominance of the applied frequency in the 
$H^{-}$ and $H^{+}$ is lost, as seen when going back from
Fig.\,\ref{heli27A} to Fig.\,\ref{heli0A}.
It is important to note that this type of helicity synchronization is 
accomplished with only minor energetic effort, as illustrated in 
Fig.\,\ref{uquadrat}. 
Synchronization occurs already (at \unit[21.5]{A}) with a
14 percent increase of the energy, and 
even at the highest forcing of \unit[27.6]{A}, the energy 
changes only by 22 percent.

As elucidated above, the sharp peak at \unit[18]{mHz} is proportional to the 
forcing while being perfectly canceled for the whole volume. It is probably directly 
driven by the forcing itself. 
What we are left with at this point is the question about the character of the 
second frequency which completely dominates the full helicity $H$.
This wider frequency bunch still shows natural variation around a center value, suggesting itself as a 
higher order mode of the slightly more irregular LSC.
As is clear from the l.h.s. of Fig.\,\ref{rolle} a non-vanishing total $H$ is
connected with a typical sloshing motion of the LSC with its 
$u_x$ that is {\it asymmetric} in $x$. This was clearly 
the dominant part for the pure RB case, 
but evidently some part of it remains also for higher forcing.
A natural explanation of the frequency increase, as observed 
in Fig.\,\ref{midfrequencies}, would be
that the spatial 
range of the azimuthal LSC's angle oscillation is restricted by the $m=2$ forcing
which dominates the $x$-region farther away from the center.
A stronger and more defined helicity oscillation would, 
in that train of thought, originate from a ``stiffer'' forcing potential wall. Just as an analogous guitar string would exhibit an increase of eigenfrequency under tension.
What is still unexplained then is the sharpness of this second peak as seen 
in Fig.\,\ref{heli12A} to Fig.\,\ref{heli27A}
which makes it tempting to speculate about a resonance 
of higher order (e.g. 4:3 or 3:2), Yet, until now 
the frequencies shown in Fig.\,\ref{midfrequencies} are not really
conclusive, so that further work on that issue is definitely required.

\section{Conclusions}

In this paper we have presented and compared experimental and numerical results 
on the synchronization of the helicity in a liquid metal RB convection 
by a time-periodic tide-like force with its typical $m=2$ azimuthal
dependence. By increasing the current in the field 
producing coils, between 10\,A and 25\,A we have observed a tightening
synchronization of the flow by the force. 
We characterized the typical periods at various 
locations in the cylinder 
and found periodic flow structures accordant to the forcing frequency.
Surprisingly, we observed also higher frequencies connected with
a remaining sidewise motion perpendicular to the
main flow direction.
We quantified the tightness of synchronicity using a 
cross correlation method 
and found strong coherence above a certain threshold.
Digging more deeply into the numerical data, 
helicity synchronization was clearly identified.
Above the threshold, the two partial 
helicities $H^{-}$ and $H^{+}$,
left and right of the LSC's main plane, showed
a nearly perfect synchronization with the external frequency,
while the full-volume helicity $H$ turned out to 
have a higher frequency, perhaps related to an 
odd-numbered synchronization (depending on peak current amplitude), an 
effect that is still to be 
understood. For different current strengths, we have visualized 
the flow which reveals a complicated three-dimensional
sloshing motion which acquires, at least partly, also a
component that is symmetric with respect to $x$.

Admittedly, the particular final results was slightly against our 
initial expectation of a rather simple synchronization of the 
side-wise sloshing motion by the $m=2$ force, allegedly leading to an 
1:1 synchronization of the
total helicity, too. Such a tidal synchronization of the
$m=0$ component of the helicity had been numerically 
observed for the Tayler instability 
\cite{Weber2015,Stefani2016}.
The difference to the RB case presented here, with its LSC in form of 
one single roll, might have to do
with the different number of $m=1$ rolls, which was equal to 2 
in the TI case (using a slightly taller cylinder).
It therefore suggests itself to use also a taller RB flow, with 
a higher number of rolls stacked above each other \cite{Schindler2022},
to better mimick the synchronization of the full-volume helicity
as in the TI case\cite{Weber2015,Stefani2016}.

Coming back to the motivating subject of a possible
synchronization of the helicity in the solar tachocline
by planetary tidal forces, it remains to be seen whether
the 1:1 synchronized {\it non-axisymmetric} helicity component, as observed 
now, can in any descent way be applied there. At any rate, it would clearly 
contradict the original idea of a Tayler-Spruit dynamo with 
{\it axisymmetric} helicity distribution. While clearly admitting that the 
analogy of our RB model with the TI in a thin tachocline 
should not be overstretched, we also point out that 
an asymmetric helicity distribution (although with respect to the equator) 
was at the root of our 1D-synchronization model \cite{Stefani2019}
of the solar dynamo.
Another interesting point might be whether any higher order
synchronization, could be helpful for explaining
some corresponding periodicities \cite{Paula2022} 
of the solar dynamo.

\section*{Acknowledgments}

This work was supported in frame of the Helmholtz - RSF
Joint Research Group ''Magnetohydrodynamic instabilities''
(contract numbers HRSF-0044 and 18-41-06201), and by the European Research 
Council (ERC) under the
European Union's Horizon 2020 research and innovation program
(grant agreement No 787544).
The results presented in this paper are based on work performed before 
February 24th 2022, 
and the research funding on the Russian side for their part 
in 18-41-06201 was terminated by the end of 2020.

We thank Dr. Tobias Vogt for his input and advice on the Rayleigh-Bénard flow and J\"org Schumacher for stimulating discussions.

\section*{Conflict of Interest}
The authors have no conflicts to disclose.

\section*{Data Availability}
The data that support the findings of this study are available from the corresponding author upon reasonable request.

\section*{Author Contributions}
\textbf{Peter J\"ustel}: Conceptualization (support), Investigation (equal), Data curation (lead), Formal analysis (lead), Methodology (equal), Visualization (lead), Software (equal), Writing - original draft (equal)
\textbf{Sebastian R\"ohrborn}: Conceptualization (support), Investigation (equal), Data curation (equal), Formal analysis (equal), Methodology (equal), Visualization (equal), Software (equal), Writing - review \& editing (equal)
\textbf{Sven Eckert}: Supervision (supporting), Methodology (supporting), Writing - review \& editing (supporting)
\textbf{Vladimir Galindo}: Visualization (supporting), Data Curation (supporting), Formal analysis (supporting), Software (equal), Methodology (supporting), Writing - review \& editing (equal)
\textbf{Thomas Gundrum}: Investigation (supporting), Software (supporting), Methodology (supporting), Resources, Writing - review \& editing (supporting)
\textbf{Rodion Stepanov}: Conceptualization (equal), Formal analysis (supporting)
\textbf{Frank Stefani}: Conceptualization (lead), Funding acquisition, Methodology (lead), Project administration, Supervision (lead), Visualization (supporting), Writing - original draft (equal)

\section*{Appendix} \label{appendix}

\begin{figure}
	\centering
	\includegraphics[width=0.48\textwidth]{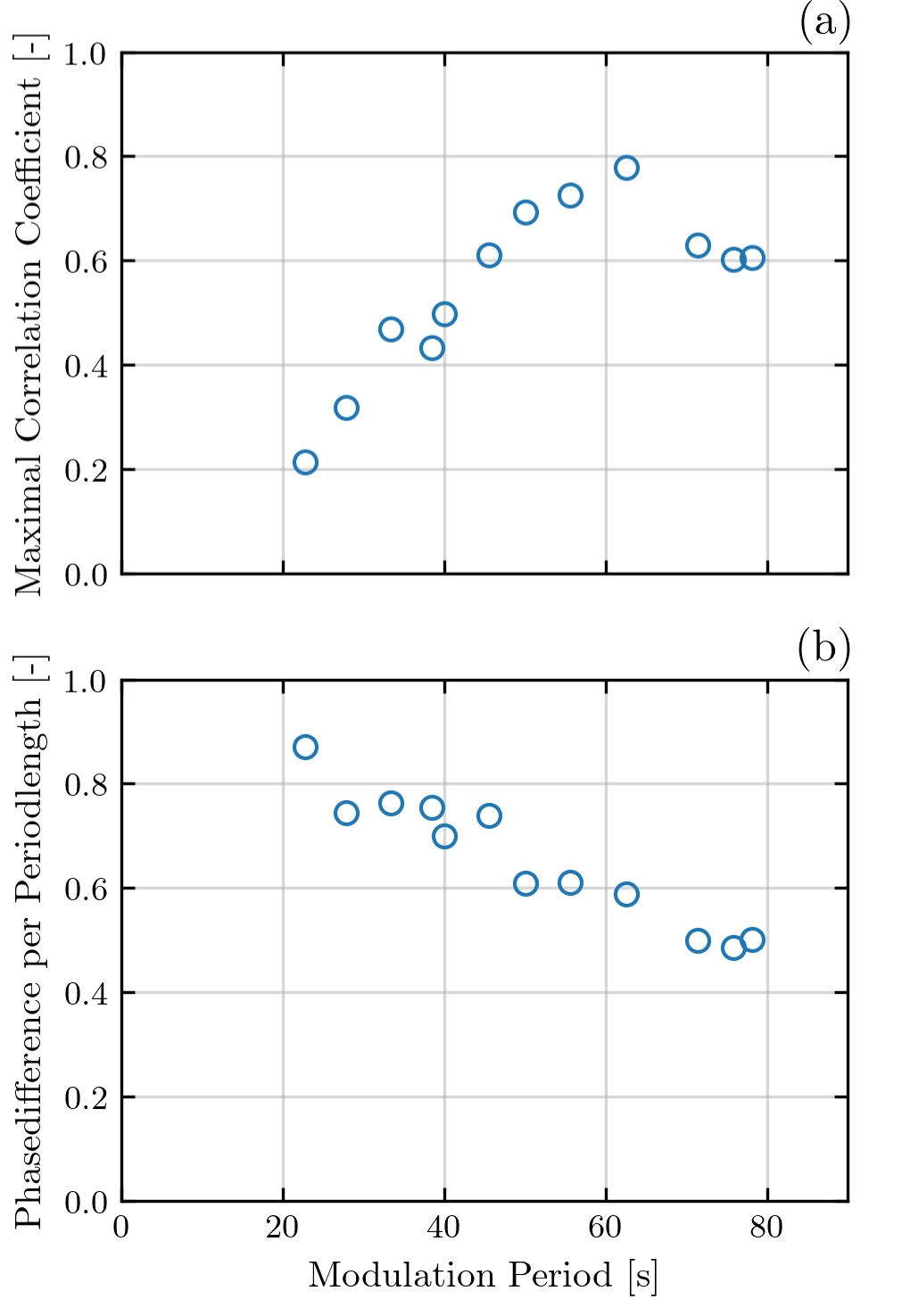}
	\caption{Maximal empirical correlation coefficients of 
		sensor ``Bot3'' as in Fig.\,\ref{correlationBoundary} for a fixed maximal current amplitude of \unit[21.2]{A} and a variation of the modulation frequency.
		(a) Correlation coefficients (b) Phase difference normalized to the modulation period.
		\label{correlationFrequencyshiftBot}}
\end{figure}

\begin{figure}
	\centering
	\includegraphics[width=0.48\textwidth]{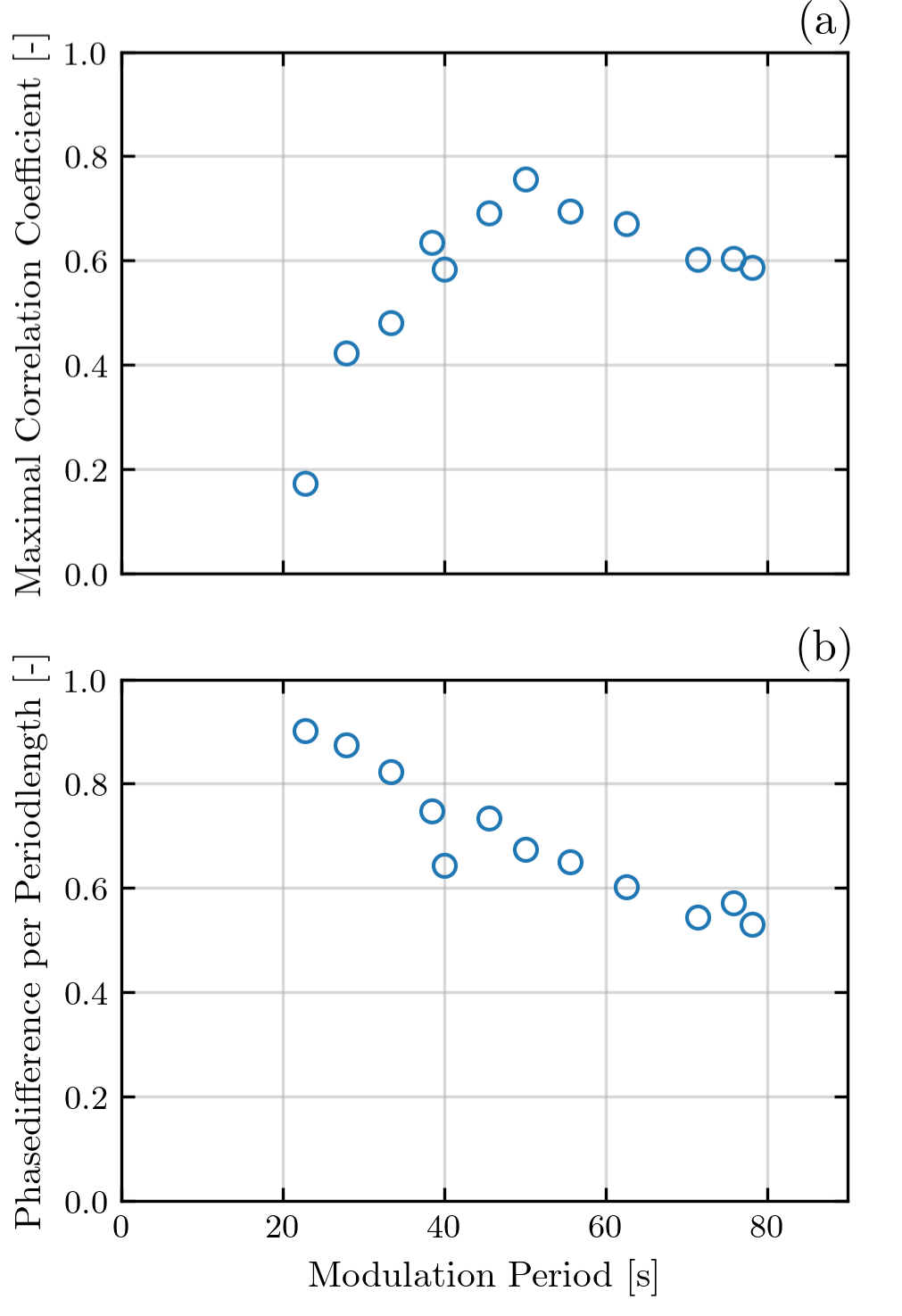}
	\caption{Same as Fig.\,\ref{correlationFrequencyshiftBot} for ``Top3'' sensor.\label{correlationFrequencyshiftTop}}
\end{figure}

Additionally to the synchronization boundary, a shift in modulation frequencies has been investigated. In Fig.\,\ref{correlationFrequencyshiftBot} the result is displayed for the ``Bot3'' sensor. It shows a broad maximum around the LSC's sloshing period of roughly \unit[55]{s}. The maximum appears to be at over \unit[60]{s}, but one has to consider the statistical variance of the values at \unit[21.2]{A}, as in Fig.\,\ref{correlationBoundary}. This is also emphasized by Fig.\,\ref{correlationFrequencyshiftTop}, where the values for the ``Top3'' sensor are given. In this case, the maximum is slightly left of \unit[55]{s}. We interpret these results such that the ``resonance'' of the forcing with the RB process is strongest at the natural sloshing frequency

The phase difference in panel (b) of Figs.\,\ref{correlationFrequencyshiftBot} and \ref{correlationFrequencyshiftTop} has been normalized to the length of one period to make the results comparable. Interestingly, the flow's lag decreases linearly from almost one period at fast modulations to about half a period for slow modulations irrespective of the apparent synchronization strength.


\begin{thebibliography}{0}%
\makeatletter
\providecommand \@ifxundefined [1]{%
 \@ifx{#1\undefined}
}%
\providecommand \@ifnum [1]{%
 \ifnum #1\expandafter \@firstoftwo
 \else \expandafter \@secondoftwo
 \fi
}%
\providecommand \@ifx [1]{%
 \ifx #1\expandafter \@firstoftwo
 \else \expandafter \@secondoftwo
 \fi
}%
\providecommand \natexlab [1]{#1}%
\providecommand \enquote  [1]{``#1''}%
\providecommand \bibnamefont  [1]{#1}%
\providecommand \bibfnamefont [1]{#1}%
\providecommand \citenamefont [1]{#1}%
\providecommand \href@noop [0]{\@secondoftwo}%
\providecommand \href [0]{\begingroup \@sanitize@url \@href}%
\providecommand \@href[1]{\@@startlink{#1}\@@href}%
\providecommand \@@href[1]{\endgroup#1\@@endlink}%
\providecommand \@sanitize@url [0]{\catcode `\\12\catcode `\$12\catcode
  `\&12\catcode `\#12\catcode `\^12\catcode `\_12\catcode `\%12\relax}%
\providecommand \@@startlink[1]{}%
\providecommand \@@endlink[0]{}%
\providecommand \url  [0]{\begingroup\@sanitize@url \@url }%
\providecommand \@url [1]{\endgroup\@href {#1}{\urlprefix }}%
\providecommand \urlprefix  [0]{URL }%
\providecommand \Eprint [0]{\href }%
\providecommand \doibase [0]{https://doi.org/}%
\providecommand \selectlanguage [0]{\@gobble}%
\providecommand \bibinfo  [0]{\@secondoftwo}%
\providecommand \bibfield  [0]{\@secondoftwo}%
\providecommand \translation [1]{[#1]}%
\providecommand \BibitemOpen [0]{}%
\providecommand \bibitemStop [0]{}%
\providecommand \bibitemNoStop [0]{.\EOS\space}%
\providecommand \EOS [0]{\spacefactor3000\relax}%
\providecommand \BibitemShut  [1]{\csname bibitem#1\endcsname}%
\let\auto@bib@innerbib\@empty
\end{thebibliography}%


%


\begin{thebibliography}{99}


\bibitem{Moffatt2018}
H.K. Moffatt, "Helicity," 
Comptes Rendus Mecanique  , \textbf{346}, 165 (2018).
\bibitem{Taylor1986}
J.B. Taylor, "Relaxation and magnetic reconnection in plasmas," 
Rev. Mod. Phys. , \textbf{58}, 741 (1986).
\bibitem{Rincon2019}
F. Rincon, "Dynamo theories," 
J. Plasma Phys., \textbf{84}, 205850401 (2019).
\bibitem{Tobias2021}
S. Tobias, "The turbulent dynamo," 
J. Fluid Dyn.  , \textbf{912}, P1 (2021).
\bibitem{Zamm2008}
F. Stefani, A. Gailitis, and G.Gerbeth, 
"Magnetohydrodynamic experiments on cosmic magnetic fields," 
Zeitschr. Angew. Math. Mech., \textbf{88}, 930 (2008).
\bibitem{Charboneau2020}
P. Charbonneau, "Dynamo models of the solar cycle," 
Liv. Rev. Solar Phys. , \textbf{17}, 4 (2020).
\bibitem{Dicke1978}
R.H. Dicke, "Is there a chronometer hidden deep in the Sun?"
Nature, \textbf{276}, 676 (1978).
\bibitem{Vos2004}
H. Vos, C. Br\"uchmann, A. L\"ucke, J.F.W. Negendank, 
G.H. Schleser, B. Zolitschka, "Phase stability of the solar Schwabe 
cycle in lake Holzmaar, Germany, and GISP2, Greenland, between 
10,000 and 9,000 cal. BP." In: Fischer, H., Kumke, T., Lohmann, G., Fl\"oser, G., 
Miller, H., von Storch, H., Negendank, J.F. (eds.), "The Climate in Historical 
Times: Towards a Synthesis of Holocene Proxy Data and Climate Models", 
GKSS School of Environmental Research, Springer, Berlin, 293 (2004).
\bibitem{AN2020}
F. Stefani, J. Beer, A. Giesecke, T. Gloaguen, M. Seilmayer, 
R. Stepanov, and T. Weier, 
"Phase coherence and phase jumps in the Schwabe cycle" 
Astron. Nachr. \textbf{341}, 600 
(2020).
\bibitem{Hung2007}
C.-C. Hung, "Apparent relations between solar activity and solar tides caused by the planets," 
NASA/TM-2007-214817 (2007).
\bibitem{Wilson2008}
I.R.G. Wilson, "Does a spin-orbit coupling between the Sun and 
the Jovian planets govern the solar cycle?" Publ. Astron. Soc. Aust. \textbf{25}, 85
(2008).
\bibitem{Wilson2013}
I.R.G. Wilson, "The Venus-Earth-Jupiter spin-orbit coupling 
model," Pattern Recogn. Phys. \textbf{1}, 147
(2013).
\bibitem{Scafetta2012}
N. Scafetta, "Does the Sun work as a nuclear fusion amplifier of 
planetary tidal forcing? A proposal for a physical mechanism based 
on the mass-luminosity relation.," J. Atmos. Solar-Terr. Phys. {\textbf 81-82}, 27 (2012).
\bibitem{Weber2015}
N. Weber, V. Galindo, F. Stefani, and T. Weier, "The 
Tayler instability at low magnetic Prandtl numbers: between 
chiral symmetry breaking and helicity
oscillations," New Journal of Physics, \textbf{17}, 113013 (2015).
\bibitem{Stefani2016}
F. Stefani, A. Giesecke, N. Weber, and T. Weier, 
"Synchronized helicity oscillations: a link between planetary tides and
 the solar cycle?" Solar Phys. \textbf{291}, 2197 (2016).
 \bibitem{Stefani2018} 
F. Stefani, A. Giesecke, and T. Weier, "On the synchronizability
of Tayler-Spruit and Babcock-Leighton type dynamos," Sol. Phys. 
\textbf{293}, 12 (2018).
\bibitem{Stefani2019} 
F. Stefani, A. Giesecke, and T. Weier, "A model of a tidally synchronized solar
dynamo," Sol. Phys. 
\textbf{294}, 60 (2019).
\bibitem{Stefani2020} 
F. Stefani, A. Giesecke, M. Seilmayer, R. Stepanov, 
and T. Weier, "Schwabe, Gleissberg, Suess-de Vries: 
Towards a consistent model of planetary synchronization of solar cycles," 
Magnetohydrodynamics \textbf{56}, 269 (2020).
\bibitem{Stefani2021} 
F. Stefani, R. Stepanov, 
and T. Weier, "Shaken and stirred: When Bond meets 
Suess-de Vries and Gnevyshev-Ohl," 
Solar Phys. \textbf{296}, 88 (2021).
\bibitem{Tayler1973}
R.~J. Tayler, "The adiabatic stability of stars 
containing magnetic fields - I.Toroidal fields," 
Mon. Not. R. Astron. Soc. \textbf{161}, 365--380 (1973).
\bibitem{Seilmayer2012}
M. Seilmayer, F. Stefani, T. Gundrum, 
T. Weier, G. Gerbeth, M. Gellert, and G. R\"udiger, 
"Experimental evidence for a 
transient Tayler instability in a cylindrical liquid-metal column,"
Phys. Rev. Lett. \textbf{108}, 244501 (2012).
\bibitem{Callebaut2012}
D.K. Callebaut, C. de Jager, S. Duhau, "The influence of planetary attractions on the
solar tachocline," J. Atmos. Sol.-Terr. Phys. \textbf{80}, 73 (2012).
\bibitem{Charbonneau2022}
P. Charbonneau, "External forcing of the solar dynamo", Front. Astron. Space Sci.
\textbf{9}, 853676 (2022).
\bibitem{Dikpati2017}
M. Dikpati, P.S. Cally, S.W. McIntosh, and E. Heifetz, 
"The origin of the "seasons" in space weather,"
Sci. Rep. \textbf{7}, 14750 (2017).
\bibitem{McIntosh2017}
S.W. McIntosh, W.J. Cramer, M. Pichardo Marcano, and R.J. Leamon, 
"The detection of Rossby-like waves on the Sun,"
Nature Astron. \textbf{11}, 0086 (2017).
\bibitem{Tobias2017}
X. M{\'a}rquez-Artavia, C.A. Jones, and S.M. Tobias, 
"Rotating magnetic shallow water waves and instabilities in a sphere,"
Geophys. Astrophys. Fluid Dyn. \textbf{111}, 282 (2017).
\bibitem{Zaqarashvili2018}
T. Zaqarashvili, 
"Equatorial magnetohydrodynamic shallow water waves in the solar tachocline,"
Astron Astrophys. \textbf{856}, 32 (2018).
\bibitem{Zaqarashvili2021}
T. Zaqarashvili et al., 
"Rossby waves in astrophysics,"
Space Sci. Rev. \textbf{217}, 15 (2021).
\bibitem{Horstmann2022}
G.Horstmann, personal communication (2022)
\bibitem{Schumacher2020}
J. Schumacher and K.R. Sreenivasan,
"Colloquium: Unusual dynamics of convection in the Sun,"
Rev. Mod. Phys. \textbf{92} 041001,  (2020)
\bibitem{Krishnamurti1981}
R. Krishnamurti, and L.N. Howard, 
"Large-scale flow generation in turbulent convection,"
Proc. Natl. Acad. Sci. \textbf{78}, 1991 (1981).
\bibitem{Sano1989}
M. Sano, X.-Z. Wu, and A. Libchaber,
"Turbulence in helium-gas free convection,"
Phys. Rev. A \textbf{40}, 6421 (1989).
\bibitem{Takeshita1996}
T. Takeshita, T. Segawa, J.A. Glazier, and M. Sano,
"Thermal turbulence in mercury,"
Phys. Rev. Lett. \textbf{76}, 1465 (1996).
\bibitem{Cioni1997}
S. Cioni, S. Ciliberto, and J. Sommeria,
"Strongly turbulent Rayleigh-B\'enard convection
in mercury: comparison with results at moderate 
Prandtl number," J. Fluid Mech. \textbf{335}, 111 (1997).
\bibitem{XiLamXia2004}
H.-D. Xi, S. Lam., and K.-Q. Xia, 
"From laminar plumes to organized flows: the 
onset of large-scale circulation in turbulent 
thermal convection," 
J. Fluid Mech. \textbf{503}, 47 (2004).
\bibitem{BrownAhlers2006}
E. Brown and G. Ahlers, 
"Rotations and cessations of the large-scale 
circulation in turbulent Rayleigh-B\'enard convection,"
J. Fluid Mech. \textbf{568}, 351 (2006).
\bibitem{Resagk2006}
C. Resagk, R. du Puits, A. Thess, F.V. Dolzhansky, S. Grossman,
F.F. Araujo, and D. Lohse, "Oscillations of the large scale wind in turbulent
thermal convection," Phys. Fluids  \textbf{18}, 095105 (2006).
\bibitem{Xi2008}
H.-D. Xi, and K.-Q. Xia, 
"Cessations and reversals of the large-scale 
circulation in turbulent thermal convection," 
Phys. Rev. E  \textbf{75}, 066307 (2008). 
\bibitem{Kadanoff2001}
L.P. Kadanoff, "Turbulent heat flow: structure and scaling," 
Phys. Today \textbf{54}, 34 (2001).
\bibitem{Funfschilling2004}
D. Funfschilling and G. Ahlers, "Plume
motion and large scale circulation in a cylindrical
Rayleigh-B\'enard cell," 
Phys. Rev. Lett. \textbf{92}, 194502 (2004).
\bibitem{XiZhou2009}
H.-D. Xi, S.-Q. Zhou, T.-S. Chan, and K.-Q. Xia,
"Origin of temperature oscillation in turbulent thermal convection,"
Phys. Rev. Lett. \textbf{102}, 044503 (2009).
\bibitem{BrownAhlers2009}
E. Brown, G. Ahlers, 
"The origin of oscillations of the large-scale
circulation of turbulent Rayleigh-B\'enard convection,"
J. Fluid Mech. \textbf{638}, 383 (2009).
\bibitem{Khalilov2018} 
R. Khalilov, I. Kolesnichenko, A. Pavlinov, 
A. Mamykin, and A. Shestakov, "Thermal convection of liquid sodium 
in inclined cylinders,"
Phys. Rev. Fluids \textbf{3}, 043503 (2018).
\bibitem{Wondrak2018} 
T. Wondrak, J. Pal, F. Stefani, V. Galindo, and S. Eckert, 
"Visualization of the global flow structure in a modified 
Rayleigh-B \'enard setup using contactless inductive flow tomography,"
Flow. Meas. Instrum. \textbf{62}, 269 (2018).
\bibitem{Zuerner2019}
T. Z\"urner, F. Schindler, T. Vogt, S. Eckert, and J. Schumacher,
"Flow regimes of Rayleigh-Bénard convection in a vertical magnetic 
field,", J. Fluid Mech. \textbf{894}, A21 (2020).
\bibitem{Morize2010}
C. Morize, M. Le Bars, P. Le Gal, and A. Tilgner,
"Experimental determination of zonal winds driven by tides,", Phys. Rev. Lett.
\textbf{104}. 214501 (2010)
\bibitem{StepanovStefani2019}
R. Stepanov, F. Stefani, 
"Electromagnetic forcing of a flow with the azimuthal wave number m = 2 in cylindrical geometry", 
Magnetohydrodynamics \textbf{55}, No. 1/2, 207-214 (2019).
\bibitem{POF2020}
P. J\"ustel et al., 
"Generating a tide-like flow in a cylindrical vessel by electromagnetic forcing", 
Phys. Fluids \textbf{32}, 097105 (2020).
\bibitem{Roehrborn2022a}
S. R\"ohrborn et al., 
"Analyzing a modulated electromagnetic $m=2$ forcing
and its capability to synchronize the large scale circulation
in a Rayleigh-B\'enard cell of aspect ratio $\Gamma=1$,"
Magnetohydrodynamics \textbf{58}, 187-193 (2022).
\bibitem{Roehrborn2022b}
S. R\"ohrborn et al., 
"Numerical simulation of the tidal synchronization 
of the large-scale circulation in Rayleigh-B'enard convection
with aspect ratio 1,"
Magnetohydrodynamics, in press (2022).
\bibitem{Pal2009}
J. Pal, A. Cramer, T. Gundrum, and G. Gerbeth, "MULTIMAG - A MUL-
TIpurpose MAGnetic system for physical modelling in magnetohydrodynamics,"
Flow Meas. Instr. \textbf{20}, 241 (2009).
\bibitem{Plevachuk2015}
Y. Plevachuk, V. Sklyarchuk, N. Shevchenko, and S. Eckert, 
"Electrophysical and structure-sensitive properties of 
liquid Ga-In alloys,"
Int. J. Mater. Res. \textbf{106}, 66 (2015).
\bibitem{LSC-Till-2019}
T. Z\"urner, F. Schindler, T. Vogt, S. Eckert, J. Schumacher, 
"Combined measurement of velocity and temperature in liquid metal convection," J. Fluid Mech. \textbf{87}, 1108 (2019).
\bibitem{openfoam}
OpenCFD, {\sl OpenFOAM - The Open Source CFD Toolbox - User’s Guide}, OpenCFD Ltd., United Kingdom, 3rd ed. (2015).
\bibitem{opera2018}
Opera, {\sl Opera - Simulation Software (Brochure), \textsuperscript{\textcopyright}Dassault Systems}, (2018)
\bibitem{Schindler2022}
F. Schindler, S. Eckert, T. Z\"urner, J. Schumacher, and T. Vogt,
"Collapse of coherent large scale flow in strongly turbulent liquid metal convection,"
Phys. Rev. Lett. \textbf{128}, 164501 (2022).
\bibitem{Paula2022}
V. de Paula, J.J. Curto, and R. Oliver,
"The cyclic behaviour in the N-S asymmetry of sunspots and solar plages for
the period 1910 to 1937 using data from Ebro catalogues",
arXiv:22.02.08628 (2022).
\end{thebibliography}
\end{document}